\documentstyle[12pt]{article}

\makeatletter
\@addtoreset{equation}{section}
\makeatother

\def\half{{\textstyle{1\over2}}}

\let\a=\alpha \let\b=\beta \let\g=\gamma \let\d=\delta

\let\la=\label  
  
\def\nn{\nonumber} \def\bd{\begin{document}} \def\ed{\end{document}}
\def\ds{\documentstyle} \let\fr=\frac \let\bl=\bigl \let\br=\bigr
\let\Br=\Bigr \let\Bl=\Bigl
\let\bm=\bibitem
\let\na=\nabla
\let\pa=\partial \let\ov=\overline
\newcommand{\be}{\begin{equation}}
\newcommand{\ee}{\end{equation}}
\def\ba{\begin{array}}
\def\ea{\end{array}}
\newcommand{\ho}[1]{$\, ^{#1}$}
\newcommand{\hoch}[1]{$\, ^{#1}$}
\newcommand{\bea}{\begin{eqnarray}}
\newcommand{\eea}{\end{eqnarray}}
\newcommand{\ra}{\rightarrow}
\newcommand{\lra}{\longrightarrow}
\newcommand{\Lra}{\Leftrightarrow}
\newcommand{\ap}{\alpha^\prime}
\newcommand{\bp}{\tilde \beta^\prime}
\newcommand{\tr}{{\rm tr} }
\newcommand{\Tr}{{\rm Tr} }
\newcommand{\NP}{Nucl. Phys. }
\newcommand{\tamphys}{\it
Center for Theoretical Physics, Department of Physics\\
Texas A\&M University, College Station, Texas 77843--4242}

\begin{document}

\rightline{CTP-TAMU-62/96}
\rightline{RU96-16-B}
\rightline{SU-ITP-96/54}
\rightline{hep-th/9612015}

\vspace{24pt}

\begin{center}
{ \large {\bf Dipole Moments of Black Holes and String States}}

\vspace{24pt}

M.~J.~Duff${}^a$\footnote{
Research supported in part by NSF Grant PHY-9411543.}, 
James T.~Liu${}^b$\footnote{
Research supported in part by 
the U.~S.~Department of Energy under grant no.~DOE-91ER40651-TASKB.}
and J.~Rahmfeld${}^c$\footnote{
Research supported by NSF Grant PHY-9219345.}

\vspace{10pt}

${}^a$ {\tamphys}

\bigskip

${}^b$ {\it Department of Physics, The Rockefeller University\\
1230 York Avenue, New York, NY 10021-6399}

\bigskip

${}^c$ {\it Department of Physics, Stanford University\\
Stanford, CA 94305-4060}

\vspace{24pt}

\underline{ABSTRACT}

\end{center}

As a further test of the conjectured equivalence of string states and
extremal black holes, we compute the dipole moments of black holes
with arbitrary spin and superspin in $D=4,N=4$ supergravity coupled to
$22$ vector multiplets and compare them with the dipole moments of
states in the heterotic string on $T^6$ or the Type IIA string on
$K3 \times T^2$.  Starting from a purely bosonic black hole with Kerr
angular momentum $L$, the superpartners are generated by acting with
fermion zero modes, thus filling out the complete supermultiplet. $L$
is then identified with the superspin. On the heterotic side,
elementary states belong only to short or long multiplets, but Type
IIA elementary states can belong to intermediate multiplets as well.
We find that the black hole gyromagnetic ratios are in perfect
agreement with the string states not only for the BPS states belonging
to short multiplets but also for those belonging to intermediate
multiplets. In fact, these intermediate multiplets provide a stronger
test of the black-hole/string-state equivalence because the
gyromagnetic ratios are not determined by supersymmetry alone, in
contrast to those of the short multiplets.  We even find agreement
between the non-supersymmetric (but still extremal) black holes and
non-BPS string states belonging to long supermultiplets. In addition
to magnetic dipole moments we also find electric dipole moments even
for purely electrically charged black holes. The electric dipole
moments of the corresponding string states have not yet been
calculated directly but are consistent with heterotic/Type IIA
duality.

\vfill
\leftline{December 1996}

\newpage

\section{Introduction}

Inspired by earlier work on black holes in string theory
\cite{Susskind:1993a}, it was conjectured in \cite{Rahmfeld1} that
massive BPS string states are described by extreme black holes. By
focusing on the elementary electrically charged states in the context
of the toroidally compactified heterotic string, results from
perturbative string theory were then used to test the conjecture. The
identification was shown to be consistent with the mass and charge
assignments in \cite{Rahmfeld1} and with the spin and supermultiplet
structures in \cite{Rahmfeld3}. A remaining check concerns the
gyromagnetic ratios, which is the subject of the present
paper%
\footnote{Preliminary results from the present paper were
presented by one of us (MJD) at {\it Strings 96}, Santa Barbara, July
1996 and the {\it European Network Meeting}, University of Crete,
Heraklion, September 1996.}.
We shall again work with the
four-dimensional $N=4$ theory which may be regarded either as a
heterotic string compactified on $T^6$ or else, by virtue of
string/string duality
\cite{Khurifour,Minasian,Khuristring,Duffstrong}, as a Type IIA
string compactified on $K3 \times T^2$ \cite{Hull,Witten}.  In such an
$N=4$ theory, elementary states belong to short, intermediate or long
supermultiplets with dimensions $16(2L+1)$, $64(2L+1)$ or $256(2L+1)$,
respectively, where $L$ is the superspin. On the heterotic side,
elementary states belong only to short or long multiplets, but Type
IIA elementary states can also belong to intermediate multiplets. We
shall determine both the electric and magnetic dipole moments of black
hole solutions with arbitrary superspin and find perfect agreement
with the string states not only for the BPS states belonging to short
multiplets but also for those belonging to intermediate multiplets as
well. In fact, these intermediate multiplets provide a stronger test
of the black-hole/string-state equivalence because the gyromagnetic
ratios are not determined by supersymmetry alone, in contrast to those
of the short multiplets.

In making this check, we shall make use of the following results in
the literature: how to generate superpartners of extreme black holes
using fermionic zero modes \cite{Aichelburg}; supersymmetric sum rules
on magnetic moments \cite{Ferrara:1992,Ferrara:1992b}; $g$-factors in
heterotic string theory \cite{Russo:1995} and black hole solutions of
heterotic strings on a torus \cite{Senblack1,Jatkar:1996}. We will
also need to bear in mind the following facts which contradict some
popular beliefs:

(i) Although the classical value $g=2$ is required in QED, the
Standard Model and indeed open string theory \cite{Argyres:1989}, this
is not a universal rule. Tree level unitarity applies only in the
energy regime $M_{\rm pl} > E > m/Q$ for a particle of mass $m$ and
charge $Q$ \cite{Ferrara:1992b}. (This also clears up an old paradox
in Kaluza-Klein theory where the massive states have $g=1$
\cite{Hosoya:1984,Kaluza}.) In the heterotic string we need two
gyromagnetic ratios for left and right movers $(g_L,g_R)$. This range
is empty for graviphoton couplings and so there are no unitarity
constraints on $g_R$; and the condition $g_L=2$ is required only for
$N_L=0$ states.  They do, however, obey the sum rule
\footnote{Since
the Type $I$ and heterotic $SO(32)$ strings are related by strong/weak
coupling \cite{Witten,HullI,Dabholkar:1995a,Polchinski:1996b}, it is
tempting to suppose that this sum rule is related to the $g=2$
condition for open strings.}
$g_L+g_R=2$ \cite{Russo:1995,Senblack2}.

(ii) Members of a supermultiplet do not necessarily have the same $g$-factor
\cite{Ferrara:1992}. 

(iii) As we shall see in this paper, in extended supersymmetry, even members 
of a supermultiplet with the same spin do not necessarily have the same
$g$-factor. 

(iv) Rotating Kerr-Newman-Sen black holes do not have the same
$g$-factor as an ``electron" (or $N_L=0$ string state); the former has
$(g_L,g_R)=(0,2)$ and the latter $(g_L,g_R)=(2,0)$: the opposite way
round!

(v) The superpartner of a non-rotating black hole is not a rotating
black hole, since the former angular momentum is provided by the
fermionic hair \cite{Aichelburg,GibbonsSUSY} and the latter by the
bosonic Kerr angular momentum.

(vi) In the {\it heterotic} string, extreme black holes are not necessarily
supersymmetric, since both $M^2=Q_R{}^2/8$ and $M^2=Q_L{}^/8$ are extreme but
only the former is supersymmetric \cite{Rahmfeld3}.  Indeed, as we shall also
discuss here, it  seems likely that the identification of string states with
extreme black holes applies to non-BPS states as well.

(vi) In addition to magnetic dipole moments we also find electric
dipole moments even for purely electrically charged black holes. The electric
dipole moments of the corresponding string states have not yet been
calculated directly but are consistent with heterotic/Type IIA duality. 
 
Further dynamical evidence for the correspondence between strings and
black holes was given in \cite{KhuriMyers,Callan:1996} where
comparisons were made between the low energy scattering amplitudes.
The idea was then taken to address the microscopic origin of black
hole entropy. It was shown \cite{Senblack2,Peet:1995} that the entropy
of these extremal black holes, evaluated at the stretched horizon
\cite{Susskind:1993}, exactly matches the result expected from the
degeneracy of string states (at least in the $N_L\gg1$ limit). More
recently, this analysis was put on more solid ground by focusing on
dyonic states which have finite scalars on the horizon ({\it e.g.}~the
four- and five-dimensional Rei\ss ner-Nordstr\o m black holes). These
states do not require the stretched horizon approach and their entropy
is uniquely determined by their finite area of the horizon. However,
it is slightly more subtle to count the number of associated string
states since non-elementary excitations are involved. This problem
could be solved \cite{Strominger:1996,Johnson:1996,Horowitz:1996} by
analyzing the black holes in the Type II picture where the D-brane
technology \cite{Polchinski:1995,Polchinski:1996a} could be applied.
As is by now well known, the results are in perfect agreement, giving
further support to the conjecture that massive string states are
described by extremal black holes \cite{Rahmfeld1}.

It is well known that the low energy limit of the heterotic theory is
described by a four dimensional $N=4$ supergravity theory coupled to
22 vector multiplets.  Denoting the vector and graviphoton electric
charges by $Q_L$ and $Q_R$ respectively, the extremal electrically
charged black holes preserving supersymmetry have masses given by
\begin{equation}
M^2=\frac{1}{8}Q_R^2
\label{Bogo}
\end{equation}
(where the dilaton VEV has been set to zero).  Saturation of this
Bogomol'nyi bound ensures that these black holes are not only
supersymmetric, but also fall in the short representation of $N=4$.
On the other hand, the the elementary string mass formula
for the heterotic string is \cite{Schwarz1}
\begin{equation}
M^2=\frac{1}{8}\left(Q^2_R+2(N_R-1/2)\right)=
\frac{1}{8}\left(Q_L^2+2(N_L-1)\right).
\label{eq:hetmass}
\end{equation}
Comparison of these expressions then indicates that if extremal black holes
obeying (\ref{Bogo}) are to be identified with string states, they must
correspond to states with oscillator numbers $N_R=1/2$ and
$N_L=1+(Q^2_R-Q^2_L)/2$. Only for vanishing Kerr angular momentum, do
solutions saturating this Bogomol'nyi bound exhibit an event horizon. Strictly
speaking, therefore, the epithet {\it black holes} should be reserved only for
the $L=0$ solutions. However, in this paper we shall use the notion of
black holes in a rather loose sense; we also refer to supersymmetric
rotating states as black holes. Otherwise one might create the impression
that they are naked singularities whereas stringy corrections to the metric
are expected and might well smooth out the singularity
\cite{Dabholkar:1995,Lu:1996}.

Previous attempts at calculating gyromagnetic ratios of black holes have
typically focused on the aspects of classically rotating solutions.
Since all such black hole solutions to the low energy $N=4$ theory have been
constructed \cite{Senblack1,Cvetic,Jatkar:1996}, it is then a simple
matter of examining the asymptotic behavior of the gauge fields in order to
read off the electric and magnetic dipole moments of any given solution.
Based on this procedure, the supersymmetric electrically charged black holes
were found to have gyromagnetic ratios $(g_L,g_R)=(0,2)$
\cite{Senblack1,Dabholkar:1995} where, as in the definitions of $Q_L$ and
$Q_R$, $g_L$ and $g_R$ correspond to vector and graviphoton couplings
respectively.  This classical result may then be compared to the
gyromagnetic ratios of elementary closed string states,
\begin{equation}
g_L=\frac{\langle S_R\rangle}{\langle S_R+S_L\rangle}, \qquad
g_R=\frac{\langle S_L\rangle}{\langle S_R+S_L\rangle},
\label{eq:gyrostring}
\end{equation}
which were derived in \cite{Russo:1995}.  In general the magnetic dipole
moments are not strictly diagonal.  Thus $\langle S_{R(L)}\rangle$ denotes
the expectation value of the spin arising from the right(left)-handed sector
of the string.  Based on this comparison, it is then seen that the classical
black hole solution corresponds to the $S_L\to\infty$ limit of the
elementary heterotic string, which fits in well with the requirement that
short representations of the string have $N_R=1/2$ and hence must get
their macroscopic spin from the left side of the string
\cite{Senblack1,Dabholkar:1995}.

Furthermore, the fact that $g_R=2$ was taken as additional support for the
idea of a black hole as a fundamental object \cite{Dabholkar:1995}.  However,
as noted above, the standard argument for demanding $g=2$, based
on the requirement of tree level unitarity, need only apply in the energy
range $M_{\rm pl} > E > m/Q$ for a particle of mass $m$ and charge $Q$
\cite{Ferrara:1992b}.  Based on (\ref{eq:hetmass}), it is easy to see that
this range is essentially empty for graviphoton couplings so that there are
no unitarity constraints on $g_R$.  On the other hand, for $g_L$, we find
that unitarity demands $g_L=2$ for $N_L=0$ states.  Since these states
contain the potentially massless gauge bosons responsible for symmetry
enhancement at special points in the $T^6$ compactification, $g_L=2$ (at
least at tree level) also follows from the requirements of gauge invariance
of the spontaneously broken Yang-Mills effective action.
What this indicates is that, if anything, unitarity constraints give the
opposite situation as found for the black holes, namely $(g_L,g_R)=(2,0)$ for 
$N_L=0$ states as opposed to $(g_L,g_R)=(0,2)$ for black holes.  Thus it is
apparent that the truly microscopic string states are not well described by
{\it classically} rotating black holes, even in the limiting case when the
spin is taken to zero.

In order to understand the finite $N_L$ states, it is necessary to realize
that the superpartners of a black hole are not simply other classical
solutions with bosonic Kerr-type angular momentum differing by small
multiples of $1/2$.  Instead, the additional spin of the black hole
superpartners is provided by fermionic hair \cite{Aichelburg}, so that a
single bosonic solution along with its fermionic hair fills out a complete
(in this case short) supermultiplet.  The aim of the present work is
to give a more complete description of supersymmetric black holes from a
manifestly supersymmetric point of view, including both bosonic Kerr-like
angular momentum and fermionic zero mode spin.  In doing so, we are able to
derive the gyromagnetic and gyroelectric ratios for all states in the
black hole supermultiplet, and find complete agreement with the
elementary heterotic string result, (\ref{eq:gyrostring}), 
for all values of $N_L$ and
not just in the limiting case.  In the process we find in fact a much
stronger result for short multiplets of $N=4$ supersymmetry---namely the
gyromagnetic and gyroelectric ratios of the short multiplets are
{\it completely} determined based on supersymmetry alone.  It should be
noted that, while it is already known that supersymmetry in general puts
restrictions on the magnetic moments \cite{Ferrara:1992}, unlike in that
case, here there is no residual freedom remaining in the choice of
transition moments in the short multiplets.

The situation is very different for black holes in intermediate 
multiplets. Always dyonic in the heterotic framework, a subclass
is elementary in the Type II picture. Here we find that supersymmetry
is far less restrictive. In fact, for fixed long-range fields (mass, charge
and angular momentum) there exists a family of distinct 
supersymmetric black hole solutions \cite{Jatkar:1996}.
Interestingly enough, they have different $g$-factors in heterotic variables.
But after translation to the Type II language, all have the proper magnetic
dipole moments
of an elementary Type II string. Thus, we see that the analysis of
gyromagnetic ratios is in fact quite powerful when it comes to
intermediate multiplets. Besides the pleasing fact that there is
an equivalence between Type II strings and extremal black holes
we also learn that generically gyromagnetic ratios in intermediate
multiplets are not purely determined by supersymmetry.

In the process of using supersymmetry to derive the magnetic dipole moments
we discover the surprising fact that {\it elementary} string states can have
electric dipole moments!  That this is the case may be readily seen by
examining elementary Kaluza-Klein states of the heterotic string and
invoking string/string duality.  Since these states have $Q_L=Q_R$ and
$(g_L,g_R)=(2,0)$ in the heterotic picture, they have a magnetic dipole
moment in both the Kaluza-Klein and winding sectors (in order to give a
purely left-handed combination).  Under dualization, this state remains
elementary, but based on the duality map must now have an electric dipole
moment in the winding sector of the Type II string.
In fact, an analysis reveals that already in the
heterotic picture the fermionic members of this multiplet
also have electric dipole moment. 
The complete scenario and interplay between gyromagnetic and gyroelectric
ratios is consistent with string/string duality, agrees with the
gyromagnetic ratios obtained from string theory, and is far more subtle than
anticipated.

The paper is organized as follows.  First, we review the low energy limit
of the heterotic string compactified on a six-torus and some
crucial facts of its duality to
the Type II string compactified on $K3\times T^2$.
Since supersymmetry properties of the solutions are crucial in the analysis,
we pay particular attention to the $N=4$ supersymmetry transformations and
briefly review the representation theory of massive $N=4$ multiplets.
In short, the supermultiplets are classified by their superspin $L$ and
the number of preserved supersymmetries $q$.  Their structure was
intensively discussed in \cite{Ferrara:1981} 
and \cite{Rahmfeld3}.  Also, $L$ was shown
in \cite{Rahmfeld3} to correspond to the Kerr angular momentum of the
supersymmetric black hole state without fermionic hair. 
In section three we first discuss general 
bosonic supersymmetric black holes. We then focus on states 
preserving two supersymmetries.  
By identifying their Killing spinors and fermionic zero modes
we generate the full short supermultiplets (up to second order
in the supersymmetry variations) and read off their 
electric and magnetic dipole moments. The particular
example of a Kaluza-Klein state is analyzed in detail and
the connection to the Type II picture is made.
Then we discuss the short multiplets in more
detail. In the heterotic language, those correspond
(mainly) to purely electric (or purely magnetic) states.

In section four we emphasize some aspects
of intermediate multiplets. The results for general states are
derived and some comments on a bound state interpretation are made. 
The freedom left by supersymmetry is emphasized. We then concentrate
on states which are elementary in the Type II picture and compare
their properties with Type II string states.
A direct correspondence is then made between extremal supersymmetric black
holes (with the appropriate quantum numbers) and elementary Type II string
states, thus providing further evidence for the identification of
black holes as elementary strings.

Before we conclude this paper we discuss the, perhaps surprising, possibility
of interpreting some non-supersym\-metric but nevertheless
extremal black holes (like the one in \cite{Garfinkle:1991}) as
elementary heterotic string states with 
$N_L=1, N_R=(Q_L^2-Q_R^2+1)/2$.
It is shown that the entropy, evaluated using the stretched horizon procedure,
and the gyromagnetic ratios agree with those of
non-BPS-saturated string states.

\section{Four-dimensional $N=4$ Supersymmetry}

In this section we briefly review the low-energy action of
the heterotic string compactified on $T^6$ and its duality
to the Type IIA string compactified on $K3\times T^2$. 
We restrict ourselves to the issues relevant 
for this paper, since this duality was historically the first 
one to be found and is widely discussed \cite{Hull,Witten,Harvey,Vafa}.
Since representation
theory lies at the heart of our results, we end this section with a summary
of the massive $N=4$ representations and how they may arise from the fermion
zero-mode algebra.

\subsection{The Heterotic Picture} 
Our starting point is the heterotic string compactified on a six-torus.  The
low-energy effective action for the string at a generic point on the Narain
lattice is given by $N=4$ supergravity with a total of 28 $U(1)$ gauge
fields. However it is important to realize that, of these 28 gauge fields,
six are singled out as graviphotons in the $N=4$ theory, with different
transformation properties from the other 22.  In terms of $N=4$
supersymmetry, the graviphotons fall in the graviton multiplet
with field content
\begin{equation}
(g_{\mu\nu}, B_{\mu\nu}, A_\mu^{(R)a}, \eta, \psi_\mu^\alpha,
\lambda^\alpha)\ ,
\label{eq:sugra}
\end{equation}
where $a$ and $\alpha$ are indices in the 6 and 4 dimensional
representations respectively of the global $SU(4)\simeq SO(6)$ symmetry
group of $N=4$ supergravity.  The other 22 $U(1)$ fields instead lie in 22
vector multiplets, given by
\begin{equation}
(A_\mu^{(L)I}, \varphi^{aI}, \chi^{\alpha I})\ ,
\end{equation}
with $I$ running from 1 to 22 labeling the multiplet.

The distinction between the graviphotons and the ordinary $U(1)$ fields
also persists at the stringy level, with the six graviphotons originating
from the right (supersymmetric) side of the heterotic string and the 22
vector fields from the left.  For this reason we have used the labels $(L)$
and $(R)$ denoting the left- and right-sided world-sheet origins of the
gauge fields.  Due to the global $SU(4)$ invariance, all six graviphotons
share identical properties.  Similarly, from the supergravity point of view,
all 22 vector multiplets may also be treated alike.  Therefore the
gyromagnetic ratios (and similarly gyroelectric ratios) for the 28 $U(1)$
fields are completely specified by just two
numbers, $g_L$ for the 22 vector multiplets and $g_R$ for the 
graviphotons\cite{Senblack1}.
In the following, it will become obvious that the different transformation
properties of the $N=4$ graviton and vector multiplets gives rise to
different values for $g_L$ and $g_R$, in accord with their left- and
right-sided origin on the string world sheet.

To be more precise about the labels $(L)$ and $(R)$ on the gauge fields, we
give a brief review of the salient properties of $N=4$ supergravity.  For
supergravity coupled to 22 vector multiplets, the $6\times 22$ scalars
$\varphi^{aI}$ parametrize the coset $O(6,22)/O(6)\times O(22)$ and may be
written in terms of a vielbein, $V$, transforming as vectors of both
$O(6,22)$ and $O(6)\times O(22)$.  The vielbein is a constrained
$28\times 28$ matrix satisfying
\begin{equation}
V^{-1} = L V^T \eta\ ,
\end{equation}
where $L$ is the $O(6,22)$ metric 
\be
L=\pmatrix{0 & I_6 & 0 \cr I_6 &0 & 0 \cr 0 & 0 &-I_{22}}\ ,
\ee
and
\begin{equation}
\eta=\pmatrix{I_6&0\cr0&-I_{22}}\ ,
\end{equation}
corresponding to the split of the vielbein into right- and left-sided [{\it
i.e.} $O(6)$ and $O(22)$] components
\begin{equation}
V=\left[\matrix{V_R\cr V_L}\right]\ .
\end{equation}
Using the vielbein allows the construction of the $O(6,22)$ matrix
\begin{equation}
{\cal M}=V^TV=V_L^TV_L^{\vphantom{T}}+V_R^TV_R^{\vphantom{T}}\ ,
\end{equation}
as well as the individual components
\begin{equation}
V_L^TV_L^{\vphantom{T}}=\half({\cal M}-L) \qquad
V_R^TV_R^{\vphantom{T}}=\half({\cal M}+L)\ .
\end{equation}

Overall, there are 28 gauge fields $A_\mu$ with field strengths $F=dA$, both
transforming as a vector of $O(6,22)$.  The vielbein is then used to
translate these field strengths into left- and right-sided ones
\begin{eqnarray}
F_{\mu\nu}^{(L)}&=&V_LLF_{\mu\nu}\nonumber\\
F_{\mu\nu}^{(R)}&=&V_RLF_{\mu\nu}\ ,
\label{eq:leftrightdef}
\end{eqnarray}
corresponding to the $N=4$ vectors and graviphotons respectively. Since
the vielbein need not be constant in general, strictly speaking only the
field strengths $F_{\mu\nu}$ and not the gauge fields $A_\mu$ themselves
may be split into right- and left-sided parts. However, whenever the
vielbein approaches its asymptotic value $V^\infty$, we may equally well
refer to left- and right-sided gauge fields according to
$A_\mu^{(R,L)}\simeq V_{R,L}^\infty LA_\mu$. This is sufficient for
determining the $g$-factors of the black holes since we only need to
examine the asymptotic behavior of the fields at infinity.

In terms of the above definitions, the low-energy effective Lagrangian for
the bosonic fields may be written in the compact form \cite{Sen6}
\begin{eqnarray}
{\cal L}&=&{1\over16\pi G}\sqrt{-g}e^{-\eta}\bigl[
R-{1\over2}(\partial\eta)^2-{1\over12}e^{-2\eta}H^2
-{1\over4}e^{-\eta}F^{(R)T}_{\mu\nu}F^{(R)\mu\nu}\nonumber\\
&&\qquad\qquad +{1\over8}\Tr (\partial {\cal M}L\partial {\cal M}L)
-{1\over4}e^{-\eta} F^{(L)T}_{\mu\nu}F^{(L)\mu\nu}\bigr]
\label{eq:lagrangian}
\end{eqnarray}
(using the canonical metric).  The first
line corresponds to the supergravity fields and the second line to the
vector multiplets.  For the heterotic string the 3-form $H$ contains a
Chern-Simons correction, $H=dB+\half A^TLdA$.  Note that the left- and
right-sided field strengths may be combined to give
\begin{equation}
F^{(R)T}_{\mu\nu}F^{(R)\mu\nu}+ F^{(L)T}_{\mu\nu}F^{(L)\mu\nu}=
F_{\mu\nu}^T(L{\cal M}L)F^{\mu\nu}\ ,
\end{equation}
resulting in a more conventional form of the Lagrangian.

For non-rotating black hole solutions, the axion is not excited and may be
consistently set to zero.  However, for rotating black holes, the axion may
not simply be discarded.  Since we only examine black hole field
configurations that solve the equations of motion, we take the liberty of
dualizing the 3-form $H$ according to
\begin{equation}
H_{\mu\nu\lambda}=e^{2\eta} \epsilon_{\mu\nu\lambda\sigma}\partial^\sigma a\ ,
\end{equation}
which defines the axion field $a$.  This allows us to write the new
Lagrangian
\begin{equation}
{\cal L}'={1\over16\pi G}\sqrt{-g}\left[
R-{|\partial S|^2\over 2(S_2)^2}
+{1\over 8}\Tr(\partial {\cal M}L\partial {\cal M}L)
-{1\over4} S_2 F^T(L{\cal M}L)F+{1\over4} S_1 F^TL*F\right]\ ,
\label{eq:axionL}
\end{equation}
which gives rise to identical equations of motion.  Here the dilaton and
axion have been combined into the complex $S$ field
\begin{equation}
S=a+ie^{-\eta}
\end{equation}
The use of $S$ results in a more compact structure in
the subsequent analysis and is additionally more natural when considering the
$SL(2;Z)_S$ properties of the black hole solutions.  However nothing depends
on this dualization, and we could equally well have retained the original
3-form $H$.

The bosonic equations of motion corresponding to either form of the $N=4$
Lagrangian are given by \cite{Sen6}
\begin{eqnarray}
&&R_{\mu\nu}={{\rm Re}(\partial_\mu S\partial_\nu \overline{S})\over2(S_2)^2}
-{1\over8}\Tr (\partial_\mu {\cal M}L\partial_\nu {\cal M}L)
+{S_2\over2}(F_{\mu\alpha}^TL{\cal M}LF_\nu{}^\alpha - {1\over4} g_{\mu\nu}
F_{\alpha\beta}^TL{\cal M}LF^{\alpha\beta})\nonumber\\
&&\nabla_\mu(-S_2{\cal M}LF^{\mu\nu}+S_1*F^{\mu\nu})=0\nonumber\\
&&{\nabla^\mu\nabla_\mu S\over (S_2)^2}+i {\partial_\mu S\partial^\mu S
\over (S_2)^3}+{1\over4}F_{\mu\nu}^TL*F^{\mu\nu}
-{i\over4}F_{\mu\nu}^TL{\cal M}LF^{\mu\nu}=0\ .
\label{eq:eom}
\end{eqnarray}
These equations are invariant under both $O(6,22;Z)_T$ and $SL(2;Z)_S$
dualities.  The latter corresponds to the transformations
\begin{equation}
S\to {aS+b\over cS+d}\qquad {\cal F}_{\mu\nu}\to (cS+d){\cal F}_{\mu\nu}\ ,
\label{eq:sduality}
\end{equation}
($ad-bc=1$) where the complex field strength ${\cal F}_{\mu\nu}$ is
defined by
\begin{equation}
{\cal F}_{\mu\nu} = F_{\mu\nu}-i {\cal M}L*F_{\mu\nu}\ ,
\end{equation}
as well as non-trivial transformations on the fermionic fields.

In order to discuss the fermions and supersymmetry, we find it convenient
to use a ten-dimensional notation for the $D=4$, $N=4$ spinors.  Hence we
introduce ten-dimensional gamma matrices with tangent space indices,
$\Gamma^A$, $A=0,2,\ldots,9$ satisfying the Clifford algebra
$\{\Gamma^A,\Gamma^B\} = 2\eta^{AB}$.  In the dimensional reduction to $D=4$,
each ten-dimensional Majorana-Weyl spinor decomposes into four $D=4$
Majorana spinors and the gamma matrices split up into a set of
four-dimensional and six-dimensional gamma matrices, given by
$\gamma^\alpha$, $\alpha=0,1,2,3$ and $\Gamma^a$, $a=1, 2,\ldots,6$
respectively ($a$ labels the 6 of $SU(4)$ as before).  Curved space
indices are then introduced in the usual manner,
$\gamma^\mu=e^\mu{}_\alpha\gamma^\alpha$.
We also define $\gamma^5=i\gamma^0\gamma^1\gamma^2\gamma^3$ (tangent space
indices) so that $(\gamma^5)^2=1$.

For a purely bosonic background, the supersymmetry transformations of the
fermions are given by
\begin{eqnarray}
\delta \psi_\mu &=& [\nabla_\mu-{1\over4}i\gamma^5 {\partial_\mu S_1\over S_2}
-{1\over 8\sqrt{2}}\sqrt{S_2} F^{(R)a}_{\alpha\beta}\gamma^{\alpha\beta}
\gamma_\mu \Gamma^a+{1\over4} Q_\mu^{ab}\Gamma^{ab}]\epsilon\nonumber\\
%
\delta\lambda&=&{1\over4\sqrt{2}}[\gamma^\mu
{\partial_\mu(S_2-i\gamma^5S_1)\over S_2}-{1\over2\sqrt{2}}\sqrt{S_2}
F_{\mu\nu}^{(R)a}\gamma^{\mu\nu}\Gamma^a]\epsilon\nonumber\\
%
\delta\chi&=&{1\over\sqrt{2}}[\gamma^\mu V_L^{\vphantom{T}}
L\partial_\mu V_R^T\cdot\Gamma
-{1\over2\sqrt{2}}\sqrt{S_2}F_{\mu\nu}^{(L)}\gamma^{\mu\nu}]\epsilon\ ,
\label{eq:susyvar}
\end{eqnarray}
where $Q_\mu^{ab} = (V_R^{\vphantom{T}}L\partial_\mu V_R^T)^{ab}$ is the
composite $SO(6)$ connection.  The first two lines correspond to the
gravitino and dilatino, which are both in the graviton multiplet, while the
last line corresponds to the gaugino.  We note that there is an obvious
decoupling between the graviton and vector multiplets, with the graviphotons
contributing to the former and the remaining 22 $U(1)$ field strengths
contributing to the latter.  Finally, we also need the lowest order
supersymmetry variation of the metric and vector fields
\begin{eqnarray}
\delta e_\mu{}^\alpha&=&\half\overline{\epsilon}\gamma^\alpha\psi_\mu
\nonumber\\
\delta A_\mu&=&-\half(S_2)^{-1/2}\overline{\epsilon}[\gamma_\mu V_L^T
\chi-\sqrt{2}V_R^T\cdot\Gamma(\psi_\mu-\sqrt{2}\gamma_\mu\lambda)]\ .
\label{eq:deltadelta}
\end{eqnarray}
Although all $6+22$ gauge fields transform together, the decoupling
between graviphotons and vector multiplets is obvious in the above variation.

We now turn to the spectrum of this theory.  This may be
characterized by two central charges, $Z_1$ and $Z_2$, which
were found in \cite{Triality,Cvetic} to be
\begin{equation}
|Z_{1,2}|^2=\frac{1}{8}\left[Q_R^2+P_R^2\pm
2\left(Q_R^2P_R^2-(Q_RP_R)^2\right)^\half\right]\ ,
\label{centralone}
\end{equation}
where $Q_R,P_R$ are the electric and magnetic charges
of the six graviphotons with field strengths $F^{(R)}_{\mu\nu}$.
For states without magnetic charge, this simply reduces to 
\begin{equation}
|Z_1|^2=|Z_2|^2={1\over8}Q_R^2\ .
\label{eq:bogoelec}
\end{equation}
On the other hand, restricting to states carrying only charges in the first
two graviphotons gives
\begin{eqnarray}
|Z_1|^2&=&\frac{1}{8}[(Q_R{}^1+P_R{}^2)^2+(Q_R{}^2-P_R{}^1)^2]
\nonumber\\
|Z_2|^2&=&\frac{1}{8}[(Q_R{}^1-P_R{}^2)^2+(Q_R{}^2+P_R{}^1)^2]\ .
\label{eq:H2Bog}
\end{eqnarray}
This restriction simplifies the discussion and is useful
when comparing heterotic states with Type II states which do
not carry Ramond-Ramond charges.

The mass $M$ of a solitonic state is bounded by the
central charges through the Bogomol'nyi bound
\begin{equation}
M\geq{\rm Max}\{Z_1,Z_2\}. \la{MMax}
\end{equation}
If $M=Z_1=Z_2$ the state preserves two supersymmetries and
is a member of a short supermultiplet. One preserved supersymmetry 
is indicated by $M=Z_1>Z_2$ which puts the state into an
intermediate multiplet. It is important that in $N=4$ supergravity
the masses of supersymmetric states are protected from quantum
corrections.  If the mass is larger than both central charges then one
has a long supermultiplet, all supersymmetries are broken, and the mass
is generically not protected from corrections.  It is clear from
(\ref{eq:bogoelec}) that elementary heterotic states are either short or
long, but never intermediate.

Let us now turn to elementary states of the heterotic string.
While the analysis is well known since the early days of string theory,
we find it instructive to review a few facts.
The heterotic string has non-supersymmetric oscillations from
the left-handed sector denoted by $\tilde\a^I_{-n}$ and
$\tilde\a^\mu_{-n}$. The right-handed
sector is supersymmetric and has bosonic oscillators $\a_{-n}^a$,
$\a_{-n}^\mu$ and fermionic ones  $b_{-r}^a$ and $b_{-r}^\mu$
acting on the Neveu-Schwarz vacuum (the Ramond sector may be treated
similarly).  The masses of the elementary string excitations are given by
\begin{eqnarray}
M^2&=&{1\over8}\left[Q_R^2+2(N_R-1/2)\right] \nn \\ 
&=&{1\over8}\left[Q_L^2+2(N_L-1)\right]
\label{eq:hetmass2}
\end{eqnarray}
where the six $Q_R$ and 22 $Q_L$ are essentially 
superpositions of momentum
and winding quantum numbers in the compactified directions
(or Yang-Mills charges); they correspond precisely to the
charges discussed before. The quantities $N_R$ and $N_L$ count the number
of string oscillator excitations. 

Comparing (\ref{eq:hetmass2}) with (\ref{MMax}), one easily
realizes, at least on the basis of masses and quantum numbers,
that supersymmetric electrically charged states could potentially be
identified with elementary string states with oscillator numbers
\be
N_R={1\over2}, \qquad N_L=\frac{1}{2}(Q_L^2-Q_R^2)+1.
\ee
All such states must necessarily lie in short multiplets, while $N_R>1/2$
states fall in long multiplets.  From the heterotic point of view
no comparison can be made between properties of states in intermediate
multiplets and elementary string states.  However this is not the case
for the dual Type II string, which does contain elementary intermediate
states.  Properties of this dual theory are summarized in the next
subsection.


\subsection{Duality to the Type II String}

It is remarkable that the Type IIA string theory compactified on
$K3\times T^2$ gives an $N=4$ theory with the exact same massless field
content and moduli space as the heterotic string compactified on $T^6$
\cite{Nilsson1,Aspinwall1}.
In this case, however, the 28 $U(1)$ gauge fields arise as 24 Ramond-Ramond
fields from $K3$ compactification as well as 2 Kaluza-Klein and 2 winding
gauge fields from the $T^2$.  Of the six $N=4$ graviphotons, four are
Ramond-Ramond, and the remaining two are combinations of the $T^2$ fields.
While much evidence has been provided in support of this duality, for the
present work we only make use of the duality map given in
\cite{PorrHet,Triality}.  Details and 
conjectures based on this duality may
be found in earlier publications \cite{Hull,Witten,Harvey,Vafa}.

In order to examine the supersymmetries of the Type II picture, we note that
the central charges derived from this theory are
\begin{equation}
|\tilde Z_{1,2}|^2=\frac{1}{8}\left[\tilde{\cal Q}^2+\tilde{\cal P}^2
\pm2\left(\tilde{\cal Q}^2\tilde{\cal P}^2-(\tilde{\cal Q}
\tilde{\cal P})^2\right)^{1\over2}\right]\ ,
\end{equation}
where ${\cal Q}$ and ${\cal P}$ are the charges of the six
graviphotons expressed in the Type II picture.  For details and
notation see \cite{Triality}.  Comparing the central charges with those on
the heterotic side, (\ref{centralone}), indicates a straightforward mapping
between the gauge fields in the two pictures.  Avoiding the Ramond-Ramond
fields, we restrict ourselves to gauge fields arising from the $T^2$ part of
the compactification, and find
\begin{eqnarray}
|\tilde Z_1|^2&=&\frac{1}{8}[(\tilde Q_R{}^1+\tilde P_R{}^2)^2
+(\tilde Q_R{}^2-\tilde P_R{}^1)^2]\nonumber\\
|\tilde Z_2|^2&=&\frac{1}{8}[(\tilde Q_L{}^1-\tilde P_L{}^2)^2
+(\tilde Q_L{}^2+\tilde P_L{}^1)^2]\ .
\label{eq:A2Bog}
\end{eqnarray}

In general, the dictionary relating the heterotic and Type II fields is
quite complicated.  However, by focusing only on $T^2$ and assuming an
essentially diagonal form of the asymptotic scalar matrix, the relevant part
of the dictionary acts on the Kaluza-Klein and winding charges as
\begin{equation}
\begin{tabular}{cc}
Heterotic&Type II\\
\hline
$Q_1$&$\widetilde{Q}_1$\\
$Q_2$&$\widetilde{Q}_2$\\
$Q_3$&$\widetilde{P}_4$\\
$Q_4$&$-\widetilde{P}_3$\\
\end{tabular}
\label{eq:dualitymap}
\end{equation}
where 1,2 are Kaluza-Klein charges and 3,4 are winding charges.  As seen
here, dualization of the 3-form has the obvious effect of electric/magnetic
duality on the winding gauge fields.  Left- and right-sided charges are
defined, in both pictures, by
\begin{equation}
Q_{R,L}^i = {1\over\sqrt{2}}(Q_i\pm Q_{i+2})
\end{equation}
and so on.  Kaluza-Klein states, with only $Q_1$ and/or $Q_2$ excited, may
clearly be elementary simultaneously in both points of view.

Let us now turn also to the perturbative spectrum of the Type
II string. Since the Type II string carries supersymmetry on
both sides we find for the mass formula \cite{Triality}
\begin{eqnarray}
M^2&=&\frac{1}{8}[(\tilde Q_R{})^2
+2(\tilde N_R-1/2)]\nonumber\\
 &=&\frac{1}{8}[(\tilde Q_L{})^2
+2(\tilde N_L-1/2)]\ ,
\label{eq:IImass}
\end{eqnarray}
{}From these formulas we learn by comparison with the central charges of
(\ref{eq:A2Bog}) (with vanishing magnetic charges)
that an elementary Type II string can be in short $(\tilde N_R=\tilde
N_L=1/2)$, intermediate $(\tilde N_{R(L)}>\tilde N_{L(R)}=1/2)$
or long $(\tilde N_{R},\tilde N_{L}>1/2)$ multiplets. This will enable
us to compare the results for the gyromagnetic ratios of black holes in
intermediate multiplets, which are out of reach of perturbative
heterotic states,
with elementary string states.

\subsection{Massive Supermultiplets and $N=4$ Representation Theory}

Before studying magnetic moments and gyromagnetic ratios, one must first
understand the nature of spinning black holes.  General rotating black
holes carrying both electric and magnetic charge have been constructed
by solving the bosonic equations of motion%
\footnote{In actuality it is often sufficient to start from a known
classical background, {\it e.g.}~the Kerr solution, and use various
symmetries to generate a complete family of solutions \cite{Senblack1}.}
with vanishing background fermions
\cite{Senblack1,Cvetic,Jatkar:1996}. Some properties 
of these solutions were outlined above. 
In addition to their mass, angular momentum and 28 dimensional
electric and magnetic charge vectors,  the black holes also
have electric and magnetic dipole moments.  Due to the dichotomy between
graviphotons and vector multiplets, the dipole moments are specified by
left- and right-sided gyromagnetic and gyroelectric ratios.

In order to relate these rotating black holes to elementary string states, it
is important to realize that, since the low-energy theory is described
by $N=4$ supergravity, all black holes must fall into irreducible $N=4$
representations, whether long, intermediate or short ones.  We
recall that massive representations of supersymmetry may be labeled by
their mass $M$ and superspin $L$ (in addition to the central charges).
A complete representation is then built up by taking the $2L+1$ states of
superspin $L$ and acting on each of them with an appropriate combination
of supercharges.  For the long representation of $N=4$, all supercharges
are active, giving rise to a $(2L+1)\times 2^{2N} = (2L+1)\times (128+128)$
dimensional multiplet.  Since the supercharges are evenly distributed
between helicities $\pm\half$, we see that generically (for $L\ge N/2$) the
long representation has spins running from $L-N/2$ to $L+N/2$
({\it i.e.}~from $L-2$ to $L+2$).  For $L<N/2$ the smallest spin is always 0.
The multiplet structure is given in Table~\ref{tbl:long}.

\begin{table}[t]
\begin{center}
\begin{tabular}{|c|rrrrr|c|c|r|}
\cline{1-6}\cline{8-9}
Spin&$L=0$&$L={1\over2}$&$L=1$&$L=3/2$&$L=2$&\qquad\qquad&Spin&$L$\\
\cline{1-6}\cline{8-9}
\noalign{\vspace{1mm}}
\cline{1-6}\cline{8-9}
4&&&&&1&&$L+2$&1\\
$\frac{7}{2}$&&&&1&8&&$L+\frac{3}{2}$&8\\
3&&&1&8&28&&$L+1$&28\\
$\frac{5}{2}$&&1&8&28&56&&$L+\frac{1}{2}$&56\\
2&1&8&28&56&70&&$L$&70\\
$\frac{3}{2}$&8&28&56&70&56&&$L-\frac{1}{2}$&56\\
1&27&56&70&56&28&&$L-1$&28\\
$\frac{1}{2}$&48&69&56&28&8&&$L-\frac{3}{2}$&8\\
0&42&48&27&8&1&&$L-2$&1\\
\cline{1-6}\cline{8-9}
\end{tabular}
\end{center}
\caption{Massive $N=4$ long representations}
\label{tbl:long}
\end{table}

For states saturating a Bogomol'nyi bound, however, not all of the
supercharges are active, as some are annihilated by the specific
representation. In this case the dimensions of the representations
become much smaller. In particular, for the short $N=4$ representation
preserving half of the supersymmetries, only half of the supercharges
are active, yielding a $(2L+1)\times 2^N=(2L+1)\times(8+8)$ dimensional
multiplet. The spin content for a short multiplet ranges from $L-1$ to
$L+1$ (for $L\ge 1$) and is summarized in Table~\ref{tbl:short} along
with the exceptional cases. Finally, states preserving a quarter of the
supersymmetries (Table~\ref{tbl:intermediate}) 
fall in intermediate multiplets of dimension
$(2L+1)\times 2^{3N/2}=(2L+1)\times(32+32)$. However they are not
present in the elementary heterotic string spectrum as they must carry
non-zero magnetic charge.

\begin{table}[t]
\begin{center}
\begin{tabular}{|c|rrr|c|c|r|}
\cline{1-4}\cline{6-7}
Spin&$L=0$&$L={1\over2}$&$L=1$&\qquad\qquad&Spin&$L$\\
\cline{1-4}\cline{6-7}
\noalign{\vspace{1mm}}
\cline{1-4}\cline{6-7}
2&&&1&&$L+1$&1\\
$3\over2$&&1&4&&$L+{1\over2}$&4\\
1&1&4&6&&$L$&6\\
$1\over2$&4&6&4&&$L-{1\over2}$&4\\
0&5&4&1&&$L-1$&1\\
\cline{1-4}\cline{6-7}
\end{tabular}
\end{center}
\caption{Massive $N=4$ short representations}
\label{tbl:short}
\end{table}

\begin{table}[t]
\begin{center}
\begin{tabular}{|c|rrrr|c|c|r|}
\cline{1-5}\cline{7-8}
Spin&$L=0$&$L={1\over2}$&$L=1$&$L=3/2$&\qquad\qquad&Spin&$L$\\
\cline{1-5}\cline{7-8}
\noalign{\vspace{1mm}}
\cline{1-5}\cline{7-8}
3&&&&1&&$L+{3\over2}$&1\\
$5\over2$&&&1&6&&$L+1$&6\\
2&&1&6&15&&$L+\frac{1}{2}$&15\\
$3\over2$&1&6&15&20&&$L$&20\\
1&6&15&20&15&&$L-\frac{1}{2}$&15\\
$1\over2$&14&20&15&6&&$L-1$&6\\
0&14&14&6&1&&$L-\frac{3}{2}$&1\\
\cline{1-5}\cline{7-8}
\end{tabular}
\end{center}
\caption{Massive $N=4$ intermediate representations}
\label{tbl:intermediate}
\end{table}

Given this digression into supersymmetry representation theory, it
should now be apparent that the Kerr angular momentum of the bosonic
solutions does not correspond directly to the spin of the state, but
instead to its superspin $L$. While the bosonic solution is just a single
member of a supermultiplet, all its superpartners may be generated by
the action of the fermion zero modes, corresponding to building up a
complete representation using the supercharges%
\footnote{This shows in particular that all superpartners of the bosonic
black hole solution have non-trivial fermionic backgrounds and allows the
possibility of spinning superpartners to the extremal black holes without
naked singularities.}.
This approach was taken in Ref.~\cite{Aichelburg} to construct exact
superpartners to the bosonic solution in the context of short $N=2$
representations.  The method was also applied in
\cite{Brooks:1995,Kallosh:1996} where further properties of $N=2$ black holes
were studied. In the following, we extend this to short and
intermediate $N=4$
representations in the context of supersymmetric elementary black hole
solutions.

Previous discussions of black hole gyromagnetic ratios have generally
ignored the fermion zero modes, and have instead focused on the bosonic
solutions with a given Kerr angular momentum.  Since the different members
of a supermultiplet may have different $g$ factors%
\footnote{While the $g$ factors for different members of a supermultiplet
are in general distinct, they are however not completely independent, but
must instead satisfy general sum rules \cite{Ferrara:1992}.},
this yields an incomplete description of the gyromagnetic ratios when $N=4$
supersymmetry is taken into account.  Furthermore, since Kerr angular
momentum $L$ corresponds directly to superspin, it is not possible to
compute the $g$ factors for the different spin components of a
supermultiplet solely by varying $L$.  Consideration of the fermion zero
modes is unavoidable if one wishes to study the properties of all
superpartners to the black holes.

In order to examine a complete black hole supermultiplet, we start with a
purely bosonic solution and then build up the rest of the representation by
acting on it with the fermionic zero modes.  Denoting the purely bosonic
solution by $\Phi$, the remaining members of the supermultiplet are then
encoded in the finite transformation \cite{Aichelburg}
\begin{equation}
\Phi\to e^\delta\Phi=\Phi+\delta\Phi+\half\delta\delta\Phi+\cdots\ .
\end{equation}
Since the supersymmetry transformation $\delta$ is given in terms of Grassman
parameters, the finite transformation eventually terminates.  However, even
for the short $N=4$ representation, this has terms up to order $\delta^8$.
So unlike the case for short $N=2$ representations, it is unrealistic to
expect to derive the exact superpartners in the present case.  We thus work
only up to the second order term since that is sufficient to see the
corrections to the bosonic background. In fact, the transformation
rules confirm this claim, since fermionic transformations
always introduce a derivative of a bosonic field (or other
fermions if one was to consider higher order fermionic 
terms). Since the bosonic fields for black hole solutions
generically behave like $B\sim  a +b/r$, second order corrections to
the bosonic fields are of order $O(\frac{1}{r^2})$ which is precisely
what we need for magnetic and electric dipole moments. Higher
order supersymmetry variations induce modifications of order
$O(\frac{1}{r^4})$ or higher and hence do not modify the
results derived from second order variations.

\section{Black Holes and String States in Short Multiplets}

After summarizing a few known results concerning bosonic supersymmetric
black holes and massive supermultiplets, 
we proceed to examine the electric
and magnetic dipole moments of states in short multiplets.  It turns out
that supersymmetry alone is sufficient to determine the properties of the
short multiplets.  Nevertheless the appearance of electric dipole moments,
while necessitated by supersymmetry, is somewhat unexpected from the
elementary string point of view.  We present a careful examination of the
electric dipole moments in the case of the Kaluza-Klein black hole, which
may be studied from both the heterotic and Type II side.

\subsection{Supersymmetric Black Holes}

A general black hole solution may be given in terms of its bosonic fields,
solving the classical equations of motion, (\ref{eq:eom}).  It is then
possible to read off the properties of this solution from the asymptotic
behavior of the fields at infinity.  In order to incorporate Kerr angular
momentum it is necessary to use a parametrization of the space-time metric
that includes off-diagonal $g_{0i}$ components.  We find it convenient to
start with a vierbein of the form
\begin{equation}
e_\mu{}^\alpha=\pmatrix{e^A&0\cr e^A C_i&e^{-A} \hat e_i{}^a}\ ,
\end{equation}
so that the metric is
\begin{equation}
g_{\mu\nu}=\pmatrix{-e^{2A}&-e^{2A}C_j\cr
-e^{2A}C_j&e^{-2A}\hat g_{ij}- e^{2A}C_iC_j}\ .
\label{eq:metric}
\end{equation}
While this decomposition is completely general, the $e^A$ factors are
motivated by the correspondence to the rotating black hole metrics of
\cite{Senblack1,Jatkar:1996}.  Asymptotic behavior of this metric determines
both the mass and the angular momentum of the black hole.  Using the
relations
\begin{eqnarray}
g_{00}&\sim&-(1-\frac{2M}{r}) \nonumber\\
g_{0i}&\sim&2 \epsilon_{ijk}L_j \frac{\hat x_k}{r^2}
\end{eqnarray}
then gives
\begin{eqnarray}
\partial_i A&\sim&M {\hat x_i\over r^2}\nonumber\\
C_i&\sim&-2\epsilon_{ijk}L_j{\hat x_k\over r^2}
\end{eqnarray}
for the metric (\ref{eq:metric}), assuming $g_{\mu\nu}\to\eta_{\mu\nu}$ at
infinity.

For electric and magnetic charges, we use the definitions
\begin{eqnarray}
E_i&\equiv&F_{0i}\sim Q\frac{\hat x_i}{r^2}\nonumber\\
B_i&\equiv&\frac{1}{2}\epsilon_{ijk}F_{jk}\sim P\frac{\hat x_i}{r^2}.
\end{eqnarray}
Since these charges are governed by the leading behavior of the fields,
the asymptotic values of the moduli, $V^\infty$, are sufficient to map
between the left and right sided charges and the combined 28 $U(1)$ charges
via $Q_{L,R}=V_{L,R}^\infty LQ$.
The electric ($\vec d$) and magnetic ($\vec\mu$) dipole moments are readily
obtained from the gauge fields $A_\mu$ themselves according to
\begin{eqnarray}
A_0&\sim&d_i\frac{\hat x_i}{r^2} \nonumber\\
A_i&\sim &\epsilon_{ijk}\mu_j\frac{\hat x_k}{r^2}\ .
\end{eqnarray}

By now, the bosonic rotating black hole solutions of toroidally compactified
string theory are completely known
\cite{Senblack1,Cvetic:1995b,Cvetic,Jatkar:1995,Jatkar:1996}.
These solutions have been constructed by acting with particular $O(8,24)$
transformations on the Kerr solution.
A subclass of solutions which is general enough to investigate dyonic black
holes was studied in \cite{Jatkar:1996}.
These states are characterized by 5 parameters: $m,l,\a,\b$ and $\g$.
$m$ and $l$ correspond to the mass and the rotation parameter of the 
original Kerr solution. The three angles $\alpha$, $\beta$ and $\gamma$
are $O(8,24)$ boost parameters which generate the solution of interest.
In general, such a solution gives a non-supersymmetric black hole.
Supersymmetric black holes are obtained generically by setting
$m\rightarrow 0$ while (at least some of) the angles approach infinity.
The condition for the absence of naked singularities is
\begin{equation}
|l|\leq m
\end{equation}
which implies that supersymmetric black holes have either no 
angular momentum or admit naked singularities. As mentioned in
the introduction we do mind the presence of such singularities, since
stringy corrections
will certainly enter and change the short-range
space-time structure anyway.

Based on the asymptotic behavior of the solution, the essential properties
of the black holes are given by \cite{Jatkar:1996,Cvetic}%
\footnote{Note that some of the signs have been changed to match our
charge and dipole moment conventions.}
\begin{eqnarray}
M&=&\frac{1}{2}m (\cosh \b^2 +\cosh \a \cosh \g) \nn \\
L&=&\frac{1}{2}l m \cosh \b (\cosh \a +\cosh \g) \nn \\
Q_R^a&=&-\sqrt{2}m\sinh \g \cosh \a \d_{a,1} \nn \\
Q_L^I&=&-\sqrt{2}m\cosh \g \sinh \a \d_{I,1}\nn \\
P_R^a&=&-\sqrt{2}m\cosh \b \sinh \b \d_{a,2}\nn \\
P_L^I&=&\sqrt{2}m\cosh \b \sinh \b  \d_{I,2}\la{classbh}\\
d_R^a&=&\sqrt{2}lm\cosh \a \sinh \b \d_{a,2}\nn \\
d_L^I&=&-\sqrt{2}lm\cosh \g \sinh \b\d_{I,2} \nn \\
\mu_R^a&=&-\sqrt{2}lm\sinh \g \cosh \b \d_{a,1}\nn \\
\mu_L^I&=&-\sqrt{2}lm\sinh \a \cosh \b \d_{I,1}\ .\nn
\end{eqnarray}
Note that use of appropriate $O(6,22)$ rotations allows non-trivial
charge configurations to be generated from this solution, which only excites
two left- and two right-sided gauge fields.
Comparing the above charge and mass formulas with (\ref{centralone}),
we can extract the supersymmetric states in the appropriate limits.  To
have a purely electrically charged supersymmetric bosonic black hole,
we have to take the limit $(m\rightarrow 0; \gamma\rightarrow \infty)$ or
$(m\rightarrow 0; \a,\g\rightarrow \infty)$ with
$M$ constant. Here, we notice already the subtle issue of limits.
For example, taking  $\a=\g\rightarrow \infty$ leaves us with
the same same mass and charge configuration but different gyromagnetic
ratios compared with a two-step procedure of first sending $\gamma$ 
and then $\alpha$ to infinity%
\footnote{This result is not necessarily in conflict with the no-hair
theorem, as the gyromagnetic ratios only have meaning for rotating
solutions, which are actually naked singularities in this limit.}.
Even worse is the situation for intermediate states, which are
dyonic in this heterotic language.  Consider the states which
are elementary in the Type II language, i.e. have $Q_R=Q_L$
and $P_R=-P_L$. For those, $\b$, $\a$ and $\g$ have to diverge;
however the details of the procedure are crucial for the electric
and magnetic moments. This will prove to be important later.

\subsection{Gyromagnetic Ratios of Short Supermultiplets}

The starting point for the fermion zero mode construction is the general
rotating black hole solution.  In the absence of fermionic hair, this black
hole is completely specified by its mass $M$ and Kerr angular momentum
$\vec L$, read off from the asymptotic behavior of the metric.

In general, rotating black hole solutions are quite complicated.  However,
in comparing the $a=\sqrt{3}$ and $a=1$ black holes to elementary string
states, since these states preserve exactly half of the supersymmetries,
this fact alone automatically determines many of the important properties
of these black holes.  Therefore we start, not with the general equations
of motion, (\ref{eq:eom}), but instead with the Killing spinor equations
derived from the supersymmetry variations, (\ref{eq:susyvar}).  Since the
graviton and vector multiplets have different properties, we examine them
individually.  Specializing to the graviton multiplet, we see that in
order for a non-trivial background to admit a Killing spinor, the various
terms in both the gravitino and dilatino variations must balance each other
out.  As a consequence, for backgrounds preserving
partial supersymmetry, the graviphotons (and hence their charges) must be
related to the dilaton-axion field.  

For short multiplets preserving exactly half of the supersymmetries, we seek
a black hole ansatz where a chiral half of the supersymmetry transformations
parametrized by $\epsilon$ is projected out.  In order to determine the form
of this projection, we first specialize to a non-rotating electric black
hole.  In this case anticipate a solution with a diagonal metric and vanishing
axion and magnetic fields.  The dilatino variation of (\ref{eq:susyvar})
then reduces to
\begin{equation}
\delta\lambda={1\over4\sqrt{2}}\gamma^i[
{\partial_i S_2\over S_2}+{1\over\sqrt{2}}\sqrt{S_2}
E_i^{(R)a}\gamma^0\Gamma^a]\epsilon\ ,
\label{eq:nonrotdil}
\end{equation}
so that preserving half of the supersymmetries demands the relation
\begin{equation}
{1\over\sqrt{2}}\sqrt{S_2}E_i^{(R)a}=\hat n^a\sqrt{-g_{00}}
{\partial_i S_2\over S_2}\ ,
\end{equation}
in which case (\ref{eq:nonrotdil}) reduces to
\begin{equation}
\delta\lambda={1\over2\sqrt{2}}\gamma^i{\partial_i S_2\over S_2} P_{\hat n}
\epsilon\ .
\end{equation}
Since there are six graviphotons, $\hat n$ is a unit vector selecting which
combination of graviphotons is excited.  This variation vanishes for exactly
half of the supersymmetries since $P_{\hat n}$ is a projection operator,
\begin{equation}
P_{\hat n}=\half(1+\gamma^{\overline{0}}\hat n\cdot\Gamma)
\label{eq:proj}
\end{equation}
($\gamma^{\overline{0}}$ denotes the Dirac matrix with flat space index).
A complete solution is then obtained by demanding that the gravitino
(as well as gaugino) variation also vanishes under the {\it identical}
projection.  

Using the projection $P_{\hat n}$ as the basis for constructing Bogomol'nyi
saturated electric black holes, we now give the complete rotating solution.
Choosing the parametrization of the metric given in (\ref{eq:metric}) and
demanding that half of the
supersymmetry is preserved according to the projection given by
(\ref{eq:proj}) now results in the first order equations
\begin{eqnarray}
E_i^{(R)}&=&-\sqrt{2}\hat n\partial_i\left({1\over S_2}\right)\nonumber\\
B_i^{(R)}&=&\sqrt{2}\hat n\epsilon_{ijk}\partial_j
\left({C_k\over S_2}\right)\ ,
\label{eq:right}
\end{eqnarray}
for the graviphotons and
\begin{eqnarray}
E_i^{(L)}\hat n&=&-\sqrt{2}{1\over S_2}V_LL\partial_iV_R^T\nonumber\\
B_i^{(L)}&=&-\epsilon_{ijk}E_j^{(L)}C_k
\label{eq:left}
\end{eqnarray}
for the left-sided $U(1)$ fields, as well as the conditions
\begin{eqnarray}
e^{-2A}&=&S_2\nonumber\\
\partial_iS_1&=&-{\hat g_{ij}\over\sqrt{\hat g}}\epsilon_{jkl}
\partial_k C_l\nonumber\\
Q_i^{ab}&=&0\ ,
\label{eq:additional}
\end{eqnarray}
and the requirement that $\hat g_{ij}$ is a flat metric%
\footnote{For a non-rotating solution these equations are a special case of
the Killing spinor equations given  in \cite{Cvetic}
where  the more general case of dyonic BPS saturated black holes preserving
both $1/2$ and $1/4$ of the supersymmetries was considered.  
In this subsection we only focus
on states preserving exactly $1/2$ of the supersymmetries, as only they
may be identified as elementary heterotic string states.}.
For simplicity we have assumed that the dilaton VEV vanishes at infinity.
However it is easy to verify that the gyromagnetic ratios are independent of
the dilaton VEV, as both charges and dipole moments scale similarly.

Remarkably enough, the above conditions already demonstrate
the consistency of the identification of black holes and
elementary string states. In the case of $Q_R=Q_L$ the string states
are required to have $N_L=1$.  Therefore the angular momentum arising from
string oscillators on the left-hand side is essentially zero.
{}From the black hole point of view this constraint was somewhat
mysterious, although addressed in higher dimensions in
\cite{Dabholkar:1995}.  Here we see a direct four-dimensional 
origin of it.  In the case of $Q_R=Q_L$ it turns out that
the black hole solutions have a vanishing axion.  But since the axion
is tied to the angular momentum by supersymmetry, as given in
(\ref{eq:additional}), it forces $L$ to vanish!  This is a very
pleasing correspondence between both pictures.

Using the above ansatz preserving half of the supersymmetries, the
variations of the fermionic fields may be expressed as
\begin{eqnarray}
\delta\psi_0&=&{1\over2}\hat\gamma^i{\partial_i(S_2-i\gamma^5S_1)\over S_2}
(S_2)^{-1}\gamma_{\overline{0}}P_{\hat n}\epsilon\nonumber\\
\delta\psi_i&=&-{1\over2}\hat\gamma^j{\partial_j(S_2-i\gamma^5S_1)\over S_2}
(\hat\gamma_i-(S_2)^{-1}C_i\gamma_{\overline{0}})P_{\hat n}\epsilon
\nonumber\\
\delta\lambda&=&{1\over2\sqrt{2}}\hat\gamma^i
{\partial_i(S_2-i\gamma^5S_1)\over S_2} (S_2)^{-1/2}P_{\hat n}\epsilon
\nonumber\\
\delta\chi&=&-\gamma^{\overline{0}}\hat\gamma^iE_i^{(L)}(S_2)^{1/2}
P_{\hat n}\epsilon\ ,
\label{eq:killing}
\end{eqnarray}
provided $\epsilon=(S_2)^{-1/4}\epsilon_0$ with $\epsilon_0$ constant.
These equations vanish for Killing spinors $\epsilon_+$ satisfying
$P_{\hat n}\epsilon_+=0$.  The remaining spinors,
\begin{equation}
\epsilon_-\quad\hbox{such that}\quad P_{\hat n}\epsilon_-=\epsilon_-
\label{eq:zeromode}
\end{equation}
then give rise to the fermion zero modes used to generate the complete
supermultiplet.  We note that this ansatz clearly shows the difference
between the
six graviphotons (part of the graviton multiplet) versus the 22 vector
multiplets.  This will come out explicitly in the results for the left- and
right-sided gyromagnetic ratios and electric dipole moments.

It is important to realize that the conditions (\ref{eq:right}),
(\ref{eq:left}) and (\ref{eq:additional}) for preserving half of the
supersymmetries are necessary but not sufficient to ensure that the
bosonic background is indeed a good black hole solution.  Recall that in
the low-energy supergravity theory, proof of the Bogomol'nyi bound through
Nester's procedure is valid only on-shell.  Equivalently we only have an
on-shell formulation of $N=4$ supergravity.  Thus the bosonic equations of
motion, (\ref{eq:eom}), must still be  imposed on the black hole solutions.  
Nevertheless, this ansatz for electric black holes, based only on the
requirement of partially unbroken supersymmetry, is sufficient to determine
many properties of the rotating black holes without having to refer to the
equations of motion.  As a result, this indicates that supersymmetric black
holes derive many of their characteristics (including gyromagnetic and
gyroelectric ratios)
solely as a consequence of their supersymmetry.  Taking the supersymmetric
limit of the general rotating solution given above (and constructed by
Sen for the case of pure electric charges
\cite{Senblack1,Senblack2,Senblack3}), we verify that the supersymmetry
ansatz is indeed satisfied.  Similarly, in the purely electric limit,
the dyonic BPS saturated black holes of \cite{Cvetic}
preserve half of the supersymmetries and manifestly satisfy the Killing
spinor equations by construction.

Before incorporating the fermion zero modes, we first read off the
properties of the bosonic background based on the electric black hole
ansatz (\ref{eq:proj}).  From the asymptotic form of the metric,
(\ref{eq:right}) shows that the right-handed electric charge vector
satisfies
\begin{equation}
Q_R=-2\sqrt{2}M\hat n\ ,
\end{equation}
which is precisely the statement that the short multiplet saturates the
Bogomol'nyi bound, given this choice of normalization factors.  For the
magnetic field, (\ref{eq:right}) is the statement that the right-sided
magnetic dipole moment is related to the angular momentum:
\begin{eqnarray}
\vec\mu_R &=& -2\sqrt{2} \vec L \hat n\nonumber\\
&=&{Q_R\over M} \vec L\ .
\end{eqnarray}
On the other hand, (\ref{eq:left}) indicates that the left-sided electric
fields are related to the scalar components of the vector multiplets and not
to any part of the graviton multiplet.  Additionally, the left-sided
magnetic dipole moment vanishes, $\vec\mu_L=0$, as both the left-sided
electric field and the Kerr angular momentum fall off as $1/r^2$.

Defining the gyromagnetic ratio by
\begin{equation}
\vec \mu={g Q\over 2M}\vec J
\label{eq:gyrodef}
\end{equation}
gives the result
\begin{equation}
g_L = 0, \qquad g_R = 2\ ,
\end{equation}
in agreement with previous results for a black hole carrying Kerr
angular momentum \cite{Senblack1,Dabholkar:1995}.  At this point we wish
to reinforce
the notion that this result is completely general for any classically
rotating solution generated by the $N=4$ short ansatz (\ref{eq:proj}),
and is derived from supersymmetry properties alone, independent of any
specific form of the black hole solution or the equations
of motion.  In order to complete this result
for general members of the $N=4$ short representation, we must now consider
the action of the fermion zero modes on this bosonic background.

Given the bosonic solution constructed to satisfy the Killing spinor
equations, all that remains is to generate its superpartners by incorporating
the fermion zero modes.  In particular, since we wish to examine the
gyromagnetic ratios, we must study the effect of the zero modes on both
the metric (for the angular momentum) and the $U(1)$ fields (for the
electric and
magnetic dipole moments).  Since we are only interested in the first
non-trivial order for the fermions, it is sufficient to look at terms
to lowest order in the fermion fields in the supersymmetry variations.  The
relevant transformations are those of the vierbein and the vector fields,
given in (\ref{eq:deltadelta}).

The variation of $g_{0i}$ yields a quantum correction to the angular
momentum.  Since the gravitino transformation for the electric black hole
is given by (\ref{eq:killing}), we find
\begin{equation}
\delta\delta g_{0i}=
-\half\overline{\epsilon}_-\left[
{\partial_jS_2\over S_2}{\hat g_{im}\over\sqrt{\hat g}}\epsilon_{mjk}
\hat\gamma_ki\gamma^5
-C_i{\partial_jS_1\over (S_2)^2}\hat\gamma^ji\gamma^5\right](S_2)^{-1/2}
\epsilon_-
\label{eq:ddg0i}
\end{equation}
where $\epsilon_-$ is a zero mode spinor defined by (\ref{eq:zeromode}).
While this expression is valid everywhere, we are only interested in its
asymptotic form.  In this case, since $\partial_i S_1\sim1/r^3$ and $C_i\sim
1/r^2$ are subdominant, the second term in (\ref{eq:ddg0i}) may be
dropped, so the asymptotic behavior is given by
\begin{eqnarray}
\delta\delta g_{0i}&\sim&-\half \epsilon_{ijk}{\partial_j S_2\over S_2}
(\overline{\epsilon}_-\hat\gamma^ki\gamma^5\epsilon_-)\nonumber\\
&\sim&-\epsilon_{ijk}M{\cal C}^j{\hat{x}^k\over r^2}\ ,
\end{eqnarray}
where ${\cal C}^i= (\overline{\epsilon}_-\hat\gamma^ii\gamma^5\epsilon_-)$
and we have used the relation between the asymptotic value of the dilaton
and the ADM mass.  This indicates that, in addition to the Kerr angular
momentum of the original rotating black hole solution, the supersymmetry
transformations generate quantum spin in direct correspondence with their
representation theory (where we view the fermion zero modes $\epsilon_-$ as
creation and annihilation operators).  The total angular momentum thus takes
the form
\begin{equation}
\vec J=\vec L+\vec S\ ,
\end{equation}
where
\begin{equation}
\vec S=-{M\over2}\vec {\cal C}
\label{eq:spindef}
\end{equation}
is the spin generated by the fermion
zero modes and $\vec L$ is the Kerr angular momentum (corresponding to
superspin) of the bosonic solution.

For the variation of the vector fields, as given in (\ref{eq:deltadelta}),
we note that both left- and right-sided gauge fields are combined into the
28 dimensional $O(6,22)$ vector $A_\mu$.  Nevertheless, the form of $\delta
A_\mu$ shows the relation between the left-sided gauge fields and the
gauginos [transforming as a vector of $O(22)$] and the relation between the
right-sided gauge fields (graviphotons) and the gravitino and dilatino.
The magnetic dipole moment is generated from the spatial components of the
gauge fields
\begin{equation}
\delta\delta A_i=\half\overline{\epsilon}_-\left[
i\gamma^5V_L^TE_j^{(L)}{\hat g_{im}\over\sqrt{\hat g}}\epsilon_{mjk}
\hat\gamma_k-(\gamma^{\overline{0}}V_L^TE_i^{(L)}+V_R^T\cdot\Gamma
{\hat n}^T\cdot E_i^{(R)})\right](S_2)^{1/2}\epsilon_-\ .
\end{equation}
Asymptotically, as the vielbein approaches $V^\infty$, the second term
becomes pure gauge.  In this case only the first term is important, and
it is immediately apparent that it induces a zero mode correction to the
magnetic dipole moment.  Substituting in the charges at infinity, we find
\begin{equation}
\delta\delta A_i\sim \half\epsilon_{ijk}V_L^TQ_L{\cal C}^j
{\hat x^k\over r^2}\ ,
\end{equation}
so that $\delta\delta\mu_i=\half V_L^TQ_L{\cal C}_i$.  Transforming to
left- and right-handed fields according to (\ref{eq:leftrightdef}) and using
the definition of spin given by (\ref{eq:spindef}), we find
\begin{equation}
\delta\delta\vec\mu_L={Q_L\over M}\vec S
\qquad \delta\delta\vec\mu_R=0\ .
\end{equation}

Previously we have seen that the purely bosonic BPS saturated electric
black hole has a magnetic moment on the right, but not the left side.
However it is clear from supersymmetry that spinning up the
black holes with the fermion zero modes creates instead a magnetic moment
only on the left.  Putting both results together, we see that 
\begin{equation}
\vec\mu_L={Q_L\over M}\vec S
\qquad
\vec\mu_R={Q_R\over M}\vec L\ ,
\end{equation}
where the total angular momentum is given by $\vec J=\vec L+\vec S$.
We note that in discussing the complete superpartners, we must appeal to a
semi-classical argument in order to make a direct correspondence with
supersymmetry representation theory since here we are adding quantum
mechanical spin $\vec S$ to classical Kerr angular momentum $\vec L$.
In this case, we see that the magnetic moments do not in general commute
with total spin $\vec J$, leading to transition dipole moments as well as
diagonal ones.  Noting that the gyromagnetic ratios defined in
(\ref{eq:gyrodef}) apply only to diagonal magnetic moments, we pick out the
$z$-axis as a preferred direction so that the $g$ factors may be written as
\begin{equation}
g_L={2\langle S^z\rangle\over L^z+S^z}\qquad
g_R={2\langle L^z\rangle\over L^z+S^z}\ .
\label{eq:gyros}
\end{equation}
These gyromagnetic ratios are summarized in Table~\ref{tbl:gyros}.

\begin{table}[t]
\begin{center}
\begin{tabular}{l|ccc}
&\multicolumn{3}{c}{Multiplicity}\\
Spin&1&4&5\\
\hline
$L+1$&${2\over L+1}$\\
$L+{1\over2}$&&${1\over L+1/2}$\\
$L$&${2\over L(L+1)}$&&0\\
$L-{1\over2}$&&$-{1\over L+1/2}$\\
$L-1$&$-{2\over L}$
\end{tabular}
\end{center}
\caption{Left-sided gyromagnetic ratios ($g_L$) for short multiplets.  The
right-sided (graviphoton) gyromagnetic ratios are given by $g_R=2-g_L$.}
\label{tbl:gyros}
\end{table}

Although derived for electric black holes in $N=4$ supergravity,
the gyromagnetic ratios that we have found are identical to those of
elementary string states
\cite{Russo:1995,Senblack1}, even up to the identification of the Kerr
angular momentum $L$ with the left side and the supermultiplet generating
spin $S$ with the right side of the heterotic string.  Therefore this
fits nicely with the  identification of black holes with
fundamental string states.  However, it should be clear in the above
derivation that supersymmetry alone has guaranteed this result.
Thus one may argue that this has not yet provided a serious test of black
holes as string states.  In the next section, however, when considering
intermediate states, we find that supersymmetry in itself is no longer
sufficient to fully constrain the electric and magnetic dipole moments.
Hence in this case a comparison of the gyromagnetic and gyroelectric ratios
will indeed provided a meaningful test of this conjecture.

\subsection{Electric dipole moments for superpartners}

We now turn to an examination of the electric dipole moments of the
supersymmetric black holes.  For the electrically charged solution
described above, it is curious that, unlike for the magnetic dipole moments,
supersymmetry alone is insufficient to determine the electric dipole
moments of the purely bosonic state.  This may be seen by noting that
non-zero electric dipole moments may be balanced against somewhat more
complicated scalar asymptotics in Eqns.~(\ref{eq:right}) and (\ref{eq:left})
of the supersymmetry conditions.  In fact non-rotating supersymmetric black
holes have even been constructed (using a different method) that
nevertheless have an electric dipole moment \cite{Horowitz:1996a}.
The occurrence of these seemingly surprising quantities can also be
understood from a slightly different angle. 
Essentially, supersymmetry and the equations of motion
require (at least for $L=0$) $e^{-\eta}$, $e^{-2A}$  and some other fields
to be harmonic functions. The canonical choice is a one-center
solution without higher moments. However, one can also add them,
leading to electric dipole moments, without violating supersymmetry or
the equations of motion. This results in solutions which are
not spherically symmetric and which have a singularity structure worse than 
the standard solution%
\footnote{The reader might argue that the singularity issue is rather
delicate in general; for example all solutions with $L\neq 0$ admit naked
singularities. We conjecture that there are ``good'' and ``bad'' naked
singularities, depending on whether they can be removed by stringy effects
or not. Since electric dipole moments in non-rotating solutions make the
space-time more singular, it appears natural to classify those
configurations as ``bad'' ones.}.  
Hence, we will not consider those kinds of solutions here. Further evidence
for disregarding them is provided by the Kaluza-Klein black hole in the
Type II picture, which will be discussed later. The main point is that 
an electric dipole moment in the heterotic language 
would be inconsistent with Type II stringy considerations.

On the other hand, the fermion zero modes give an additional
electric dipole moment contribution to the superpartners which is completely
determined by supersymmetry.  For the zero modes we find
\begin{equation}
\delta\delta A_0={1\over2}\overline{\epsilon}_-\left[\hat n\cdot E_i^{(R)}
\hat\gamma^i\gamma_{\overline{0}}V_R^T\cdot\Gamma\right](S_2)^{-1/2}
\epsilon_-\ ,
\end{equation}
leading to a contribution to the right-sided electric dipole
moment
\begin{equation}
\delta\delta \vec d_R^a=-{1\over2}\hat n\cdot Q_R\vec{\cal D}^a\ ,
\end{equation}
where ${\cal D}^{i\,a}=(\overline{\epsilon}_-\gamma_{\overline{0}}
\hat\gamma^i\Gamma^a\epsilon_-)$.

Unlike for the zero-mode-induced magnetic dipole moment,
$\delta\delta\vec\mu_L$, the resulting electric dipole moment is not
proportional to the zero-mode spin $\vec S$.  Furthermore, an electric
dipole moment is generated for potentially all six graviphotons, even
though the electric charge is carried by a single combination
proportional to $\hat n$.  Noting that $\hat n\cdot \vec {\cal D}=0$,
it is in fact only the five graviphotons {\it orthogonal} to $\hat n$
that couple to the electric dipole moment.  In other words, when an
electric charge is turned on under a given graviphoton, it never picks
up an electric dipole moment, while the other five do.

In general, since $\vec{\cal D}^a$ and $\vec{\cal C}$ have a different
structure, the zero-mode electric dipole operator does not commute
with spin, leading to transition moments and the impossibility of
diagonalizing all five graviphoton electric dipole moments simultaneously.
For a particular graviphoton (orthogonal to $\vec n$), it is possible to
see that $\vec{\cal D}=\pm\vec{\cal C}$, where the sign depends on the
chirality of the zero mode spinor $\epsilon_-$ in a specific
six-dimensional subspace of the original ten-dimensional space-time.  In
particular, it is instructive to consider the $T^2$ truncation of the
toroidally compactified heterotic string.  Picking $\hat n^a=\delta^{a1}$
then gives
\begin{eqnarray}
\delta\delta\vec d_R^1&=&0\nonumber\\
\delta\delta\vec d_R^2&=&\pm{Q_R\over M}\vec S\ ,
\label{eq:edmcase}
\end{eqnarray}
keeping in mind that $\delta\delta\vec d_R^a$ for $a=3,4,5,6$ are
non-vanishing (and not diagonal) in this basis.

Since supersymmetry alone does not fix the electric dipole moments of a
short black hole, we now turn to the explicit solution of \cite{Senblack1}.
This rotating electric black hole corresponds to setting $\beta=0$ in the
more general dyonic solution with charges given by Eqn.~(\ref{classbh}).
In this case it is easy to see that the bosonic state has vanishing electric
dipole moments.  Furthermore, examination of (\ref{eq:edmcase}) indicates
that only the four spin $L+1/2$ and the four spin $L-1/2$ members of the
superspin $L$ short multiplet have electric dipole moments (if we start out
with a solution without an electric dipole moment).  This also
demonstrates that even potentially light states, such as the spin 1/2
``gauginos'' of the $L=0$ multiplet have electric dipole
moments.  While on the surface this appears troublesome for low-energy
phenomenology, it should be noted that such electric dipole moments appear
only in the graviphoton couplings, and not to the ordinary $U(1)$ vector
fields.

%
%
%
%

In retrospect, the appearance of electric dipole moments is not really
surprising from the point of view of heterotic/Type II duality.  Focusing
on $T^2$,
we consider an electric Kaluza-Klein state which may be elementary in
both pictures.  From the heterotic point of view we let this state carry
electric charge $Q_1=q$ ({\it i.e.}~$Q=(q,0,0,0)$ where the first two
components are Kaluza-Klein charges and the last two are winding charges),
so that $Q_L^1=Q_R^1=q/\sqrt{2}$.  Focusing on the $L=0$ case (since $N_L=1$,
the only possibilities are $L=0$ or $L=1$), we have seen that the heterotic
gyromagnetic ratios are given by $(g_L,g_R)=(2,0)$.  In turn, this implies
that $(\vec\mu_L^1,\vec\mu_R^1) =(-2\sqrt{2}\vec J,0)$, or, in terms of
Kaluza-Klein and winding fields, $\vec\mu=(-2\vec J,0,2\vec J,0)$.  Using
the duality map of (\ref{eq:dualitymap}), this translates into the Type II
side as
\begin{eqnarray}
\vec{\tilde\mu}_1&=&-2\vec J\nonumber\\
\vec{\tilde d}_4&=&-2\vec J\ ,
\end{eqnarray}
so that, starting only with magnetic dipole moments, we are inevitably led
to consider electric dipole moments.

To complete the picture we may also use the duality map in reverse, and
start with an elementary Type II string.  Since the stringy formula,
(\ref{eq:gyrostring}), is in fact applicable to {\it all} closed strings, it
enables us to determine the gyromagnetic ratios of the Kaluza-Klein state
in the Type II picture.  Based on an orbifold construction of the $K3$
compactification, the short superspin 0 multiplet is generated by the
left/right spin combination $[(1/2)+2(0)]_L\times[(1/2)+2(0)]_R =
[(1)+4(1/2)+5(0)]$, and has gyromagnetic ratios
\begin{equation}
(\tilde g_L,\tilde g_R)=\left[\matrix{(1,1)\cr
2\times (2,0),\ 2\times (0,2)\cr
5\times (\cdot,\cdot)}\right]\ ,
\end{equation}
in agreement with the $L=0$ Kaluza-Klein black hole results given in
Table~\ref{tbl:kkbh}%
\footnote{Note that in particular one finds vanishing magnetic dipole
moments for the $S=0$ members, which implies that the electric dipole
moments should vanish in the heterotic picture, as discussed before.}.
Translating these magnetic moments to the heterotic picture then requires
that electric dipole moments are present in the heterotic side as well.

\begin{table}[t]
\begin{center}
\begin{tabular}{c||c|cccc|cccc}
\multicolumn{2}{c}{}&
\multicolumn{4}{c}{Heterotic}&\multicolumn{4}{c}{Type II}\\
spin&&$F_1$&$F_2$&$F_3$&$F_4$&$\tilde F_1$&$\tilde F_2$&
$\tilde F_3$&$\tilde F_4$\\
\hline
\hline
&$Q$&$q$&0&0&0&$q$&0&0&0\\
\hline
(1)&$\vec\mu$&$-2\vec J$&0&$2\vec J$&0&$-2\vec J$&0&0&0\\
&$\vec d$&0&0&0&0&0&0&0&$-2\vec J$\\
\hline
$2\times(1/2)$&$\vec\mu$&$-2\vec J$&0&$2\vec J$&0&$-2\vec J$&0&$2\vec J$&0\\
&$\vec d$&0&$-2\vec J$&0&$-2\vec J$&0&$-2\vec J$&0&$-2\vec J$\\
\hline
$2\times(1/2)$&$\vec\mu$&$-2\vec J$&0&$2\vec J$&0&$-2\vec J$&0&$-2\vec J$&0\\
&$\vec d$&0&$2\vec J$&0&$2\vec J$&0&$2\vec J$&0&$-2\vec J$\\
\hline
$5\times(0)$&$\vec\mu$&0&0&0&0&0&0&0&0\\
&$\vec d$&0&0&0&0&0&0&0&0\\
\end{tabular}
\end{center}
\caption{Electric charges and magnetic and electric dipole moments of the
$L=0$ Kaluza-Klein black hole supermultiplet in both the heterotic and Type
II pictures.}
\label{tbl:kkbh}
\end{table}

What we have basically shown is that supersymmetry itself is sufficient
to require the existence of electric dipole moments in the graviphoton
couplings of short superspin $L$ multiplets (although it does not
necessarily fix their values, as demonstrated in \cite{Horowitz:1996a}).
It remains an open issue how
these electric dipole moments originate from the string world sheet point of
view.  Since such electric dipole moments are presumably not present in the
bosonic string, we conjecture that they arise as a consequence of
world-sheet supersymmetry, and perhaps only in the Ramond sector of the
superstring.  This picture is of course consistent with the states shown in
Table~\ref{tbl:kkbh}.

\section{Intermediate Multiplets and Gyromagnetic Ratios}

While we have seen that supersymmetry leaves no freedom for the electric and
dipole moments of states in short multiplets, it turns out that the states
in intermediate multiplets are far less constrained.  By using a combination
of ``electric'' and ``magnetic'' projection operators in the Killing spinor
equations, we are able to extend the short multiplet black hole ansatz of
the previous section to cover the case of intermediate multiplets.  In doing
so, we find the intriguing picture that a dyonic black hole may essentially
be viewed as a combination of separate electrically and magnetically charged
black holes.

Although the heterotic string has no elementary states in intermediate
multiplets, we may use the duality map to go over to the Type II picture
where such states do exist.  Thus we make a
comparison of certain intermediate black holes with elementary Type II
string states.  The resulting picture is considerably more complicated than
that for short multiplets.  At the end of this section we also make a few
comments on how this analysis may be extended to consider non-supersymmetric
states.

\subsection{Supersymmetry and the intermediate multiplet solution}

We start by recalling that many properties of the basic supersymmetric
electric black hole solution may be determined starting from the electric
projection operator $P_{\hat n}$ of Eqn.~(\ref{eq:proj}).  Since the
$N=4$ supersymmetries and equations of motion are $S$-duality invariant,
it is clear that dyonic black holes in {\it short} multiplets may
similarly be constructed by means of a duality rotated projection,
\begin{equation}
P_{\hat n}^\theta={1\over2}(1+e^{i\gamma^5\theta}\gamma^{\overline{0}}
\hat n\cdot\Gamma)\ .
\end{equation}
In particular, choosing $\theta=\pi/2$ corresponds to a magnetic projection,
and leads to short magnetic black holes.

Intermediate states, on the other hand, may be constructed using a
combination of the electric projection operator $P_{\hat n}$ and a
magnetic projection
\begin{equation}
\tilde P_{\hat m}= \frac{1}{2}(1+i\g^5\g^{\bar 0}\hat m \cdot \Gamma)\ .
\end{equation}
We note that such intermediate states are always dyonic in the heterotic
language since no duality rotation may remove completely the magnetic 
charge generated by the combination of $P_{\hat n}$ and $\tilde P_{\hat m}$.

As in the previous section, construction of the intermediate solution 
starts from the general rotating metric ansatz, (\ref{eq:metric}).  However
this time we demand that only a quarter of the supersymmetries are preserved.
In this case, the expressions are more complicated and have more freedom.
The resulting graviphoton fields are given by
\begin{eqnarray}
E_i^{(R)} &=& -\sqrt{2} e^{A+\eta/2}
(\hat n \partial_i (A+\eta/2) - \hat m \partial_i X)\nonumber\\
{\hat g_{ij}\over\sqrt{\hat g}}(B^{(R)}_j+\epsilon_{jkl}E_k^{(R)}C_l)
&=& \sqrt{2} e^{-(A-\eta/2)}
(\hat m \partial_i(A-\eta/2) + \hat n\partial_i Y)\ ,
\label{eq:dyonR}
\end{eqnarray}
where $\hat n$ and $\hat m$ determine the graviphoton electric and magnetic
charge vectors respectively.  The angular momentum, specified by $C_i$, has
been split into two contributions,
\begin{equation}
e^{2A}{\hat g_{ij}\over\sqrt{\hat g}}\epsilon_{jkl}\partial_kC_l
= \partial_iX+\partial_iY\ ,
\end{equation}
where
\begin{equation}
\partial_iY\equiv\partial_iX - e^\eta\partial_ia\ .
\end{equation}
As a result, (\ref{eq:dyonR}) indicates that both electric and magnetic
graviphoton dipole moments are induced by non-vanishing Kerr angular momentum,
in proportion to possible splitting of angular momentum between $\partial_iX$
and $\partial_iY$.
At this point is is convenient to make the asymptotic definitions of masses,
\begin{eqnarray}
\partial_i(A+\eta/2)&\sim&2M_1{\hat x_i\over r^2}\nonumber\\
\partial_i(A-\eta/2)&\sim&2M_2{\hat x_i\over r^2}\ ,
\end{eqnarray}
and angular momenta,
\begin{eqnarray}
\partial_i Y&\sim&2\partial_i\left({\vec L_1\cdot\hat x\over
r^2}\right)\nonumber\\
\partial_i X&\sim&2\partial_i\left({\vec L_2\cdot\hat x\over
r^2}\right)\ ,
\end{eqnarray}
so that the intermediate black hole has mass $M=M_1+M_2$ and Kerr angular
momentum $\vec L=\vec L_1+\vec L_2$.  While this splitting may appear
{\it ad hoc}, as we will see, it has its basis in the ``splitting'' of the
dyonic black hole into individual electric and magnetic components.

It is crucial that the functions $X$ and $Y$ are determined
not only by the Kerr angular momentum (which characterizes the superspin of
the multiplet), but also by the axion, which is off-hand 
unrelated to the supersymmetry representation. The conclusion is 
that states with the same spin $L$ and the same charges could
have a different splitting of their spin into $L_1$ and $L_2$
components, which ultimately leads to different $g$-factors.  Thus the case
of intermediate multiplets is already quite different from that of short
multiplets.  In fact, this ambiguity is precisely reflected by the
limit dependence in (\ref{classbh}). The difference between $\a$ and
$\g$ does not affect the charges, the mass and the angular momentum,
as long those angles are eventually taken to be infinite.  However
the moments are very sensitive to the particular choice of how the limit
is obtained.  In short, supersymmetry allows for a wide class of
solutions with the same quantum numbers, but different dipole moments
and hence different gyromagnetic/gyroelectric ratios.
We will see later how string states tie in with this.  But it is already
apparent that supersymmetry in general no longer fixes the gyromagnetic
ratios of states in intermediate representations.

For the left-sided gauge fields, the tale is somewhat more complicated.
Since these fields belong to $N=4$ vector multiplets, the condition for
generating intermediate states now relates the scalar and vector
components of this multiplet according to
\begin{equation}
\sqrt{2} V_L^{\vphantom{T}}L\partial_iV_R^T = -e^{-(A+\eta/2)}E_i^{(L)} \hat n
-e^{A-\eta/2}{\hat g_{ij}\over\sqrt{\hat g}}(B_j^{(L)}
+\epsilon_{jkl}E^{(L)}_kC_l)\hat m\ .
\label{eq:dyonL}
\end{equation}
Due to the presence of both electric and magnetic terms, unlike for the
short ansatz, here no general statement can be made about the left-sided
electric and magnetic dipole moments.  Only after solving the equations of
motion can one examine either the asymptotics of the scalars or the
left-sided gauge fields to determine what dipole moments are generated.  So
once again we find considerably more freedom for intermediate multiplets.

With this intermediate state ansatz, the Killing spinor equations following
from (\ref{eq:susyvar}) become
\begin{eqnarray}
\delta\psi_0&=&-{1\over2}e^{2A}\hat\gamma^i\gamma_{\overline0}
[\partial_i(A+\eta/2+i\gamma^5Y)P_{\hat n}
+\partial_i(A-\eta/2+i\gamma^5X)\tilde P_{\hat m}]\epsilon\nonumber\\
\delta\psi_i&=&{1\over2}\hat\gamma^j(\hat\gamma_i-e^{2A}C_i
\gamma_{\overline0})
[\partial_j(A+\eta/2+i\gamma^5Y)P_{\hat n}
+\partial_j(A-\eta/2+i\gamma^5X)\tilde P_{\hat m}]\epsilon\nonumber\\
\delta\lambda&=&-{1\over2\sqrt{2}}e^A\hat\gamma^i
[\partial_i(A+\eta/2-i\gamma^5Y)P_{\hat n}
-\partial_i(A-\eta/2-i\gamma^5X)\tilde P_{\hat m}]\epsilon\nonumber\\
\delta\chi&=&-e^A\gamma^{\overline0}\hat\gamma^i
[e^{-(A+\eta/2)}E_i^{(L)}P_{\hat n}
-i\gamma^5e^{A-\eta/2}{\hat g_{ij}\over\sqrt{\hat g}}(B_j^{(L)}
+\epsilon_{jkl}E_k^{(L)}C_l)\tilde P_{\hat m}]\epsilon\ .
\end{eqnarray}
In addition to the gauge field ansatz, the composite $SO(6)$ connection
vanishes, $Q_i^{ab}=0$, and the spinor parameter is related to a constant
spinor $\epsilon_0$ according to $\epsilon=e^{(A+i\gamma^5X)/2}
\epsilon_0$.  In general, these expressions decompose as a sum of two sets
of conditions, one based on $P_{\hat n}$ and the other based on
$\tilde P_{\hat m}$.  In order to construct an intermediate state preserving
a quarter of the supersymmetries, it is easiest to impose 
$\hat n\cdot \hat m=0$
so that both projections commute with each other.  This condition is
essentially a no-force condition ensuring that the electric and magnetic
black hole states generated by $P_{\hat n}$ and $\tilde P_{\hat m}$ are
orthogonal.

In fact, based on this construction, it is apparent that the dyonic black
holes preserving a quarter of the supersymmetries may be viewed as a
combination of an electric and a magnetic black hole.  Generating complete
supermultiplets via the fermion zero modes, we find that the electrically
charged state has mass $M_1 = -\hat n\cdot Q_R/2\sqrt{2}$ and left- and
right-sided magnetic dipole moments given by
\begin{equation}
(\vec\mu_L,\vec\mu_R) =
 (\vec\mu_L^{(0)},0)+ 
{1\over M_1} (Q_L\vec S_1,Q_R\vec L_1)\ ,
\end{equation}
where $\vec\mu_L^{(0)}$ indicates the original left-sided magnetic moment
which is undetermined by supersymmetry%
\footnote{Since $\vec\mu_L^{(0)}$ is not fixed by supersymmetry, at this
point there is no way to assign it to either the electric or the magnetic
state.  In the next subsection, we see that in fact $\vec\mu_L^{(0)}$
belongs to the {\it magnetic} state, and not the electrically charged one.}.
As before, $\vec L_1$ and $\vec S_1$ correspond to Kerr angular momentum and
quantum spin, respectively, with
\begin{equation}
\vec S_1 = -{M_1\over2}(\overline\epsilon\vec\gamma i\gamma^5
P_{\hat n}\epsilon)\ .
\end{equation}
Additionally, the fermion zero modes generate graviphoton electric dipole
moments for the five graviphotons orthogonal to $\hat n$
\begin{equation}
\vec d_R^a=-{1\over2}\hat n\cdot Q_R(\overline\epsilon
\gamma_{\overline{0}}\vec\gamma\Gamma^aP_{\hat n}\epsilon)\ .
\end{equation}

The magnetically charged state, on the other hand, has mass $M_2=\hat m\cdot
P_R/2\sqrt{2}$ and electric dipole moments
\begin{equation}
(\vec d_L,\vec d_R)=(\vec d_L^{(0)},0)
-{1\over M_2}(P_L\vec S_2, P_R\vec L_2)\ ,
\end{equation}
this time with
\begin{equation}
\vec S_2 = -{M_2\over2}(\overline\epsilon\vec\gamma i\gamma^5
\tilde P_{\hat m}\epsilon)\ .
\end{equation}
Graviphoton magnetic dipole moments are also generated according to
\begin{equation}
\vec\mu_R^a={1\over2}\hat m\cdot P_R
(\overline\epsilon\gamma_{\overline{0}}\vec\gamma i\gamma^5\Gamma^a
\tilde P_{\hat m}\epsilon)\ .
\end{equation}
The complete dyon then gets its properties from a sum of the individual
electric and magnetic contributions. In particular, the mass $M = M_1+M_2
= (-\hat n\cdot Q_R + \hat m\cdot P_R)/2\sqrt{2}$ confirms that this
dyon saturates the Bogomol'nyi bound appropriate for an intermediate $N=4$
state.  This description of intermediate black holes is summarized in
Table~\ref{tbl:inter}.

\begin{table}[t]
\begin{center}
\begin{tabular}{l|cc}
&Electric state&Magnetic state\\
\hline
Projection&$P_{\hat n}={1\over2}(1+\gamma^{\overline{0}}\hat n\cdot\Gamma)$
&$\tilde P_{\hat m}={1\over2}
(1+i\gamma^5\gamma^{\overline{0}}\hat m\cdot\Gamma)$\\
Masses&$M_1$&$M_2$\\
Kerr angular momentum&$\vec L_1$&$\vec L_2$\\
Charges&$Q_R=-2\sqrt{2}M_1\hat n$&$P_R=2\sqrt{2}M_2\hat m$\\
&$Q_L=\ldots\kern+40pt $&$P_L=\ldots\kern+33pt$\\
Dipole moments&$\vec\mu_R=-2\sqrt{2}\vec L_1 \hat n$
&$\vec d_R=-2\sqrt{2}\vec L_2 \hat m$\\
&\multicolumn{2}{c}{$\vec\mu_L=\ldots$}\\
&\multicolumn{2}{c}{$\vec d_L=\ldots$}\\
\hline
Zero-mode spin&$\vec S_1=-{1\over2}M_1(\overline{\epsilon}\vec\gamma
i\gamma^5 P_{\hat n}\epsilon)$
&$\vec S_2=-{1\over2}M_2(\overline{\epsilon}\vec\gamma
i\gamma^5 \tilde P_{\hat m}\epsilon)$\\
Zero-mode generated&$\delta\delta\vec\mu_L={Q_L\over M_1}\vec S_1$
&$\delta\delta\vec d_L=-{P_L\over M_2}\vec S_2$\\
~~~~dipole moments&$\delta\delta\vec d_R=\pm {\hat n\cdot Q_R\over
M_1}\vec S_1$
&$\delta\delta\vec\mu_R=\pm {\hat m\cdot P_R\over M_2}\vec S_2$\\
\end{tabular}
\end{center}
\caption{Properties of intermediate states, viewed as a combination of
individual electric and magnetic states.  Note that the splitting of angular
momentum as well as some left-sided dipole moments are undetermined by
supersymmetry.  The $\pm$ sign for the zero-mode-generated graviphoton
dipole moments is only schematic, and is discussed in further detail in the
text.}
\label{tbl:inter}
\end{table}

We must emphasize that, while there is a natural decomposition of the
intermediate black hole into its ``electric'' and ``magnetic'' parts, there
is only a single set of fermion zero modes which act on both parts
simultaneously.  Thus it is not necessarily the case that we may conclude
from this ansatz that the intermediate black hole is a direct result of
combining two marginally stable short black holes.  Any true bound state
interpretation of such dyonic black holes certainly requires a more
careful analysis of this solution.

\subsection{Black Holes and Type II Strings}

It is now very interesting to compare those intermediate black hole states
with elementary string states.  Since the elementary heterotic string
admits no intermediate states, this comparison can only be made to the Type
II string, which does have elementary excitations in the intermediate
multiplets.  Since the general picture is somewhat complicated, we first
consider the large superspin limit, $L\gg1$, which corresponds to the
classically rotating black hole where only Kerr angular momentum
needs to be considered.  After examining the purely bosonic solution, we
then study the incorporation of fermion zero modes to generated the complete
intermediate multiplet.

In the $L\gg1$ limit the string mass formula, (\ref{eq:IImass}), requires
either $\tilde N_R=1/2$ or $\tilde N_L=1/2$ for states in an intermediate
multiplet.  Thus the Kerr angular momentum must come entirely from either
left- or right-sided string oscillators.  Using the string formula for
gyromagnetic ratios, (\ref{eq:gyrostring}), leads to
\be
\tilde g=(0,2) \qquad {\rm or} \qquad \tilde g=(2,0)\ ,
\la{gIILb}
\ee
for $\tilde N_R=1/2$ or $\tilde N_L=1/2$ states respectively.

This may now be compared with the bosonic black hole solutions of
\cite{Senblack1,Cvetic,Jatkar:1996}, with asymptotic behavior
given in Eqn.~(\ref{classbh}).  Intermediate dyonic states may be generated
by taking the simultaneous limit $\alpha,\beta,\gamma\to\infty$ and
$m\to0$ while keeping the following quantities fixed:
\begin{eqnarray}
M_1&=&\half m\cosh\alpha\cosh\gamma\nonumber\\
M_2&=&\half m\cosh^2\beta\nonumber\\
L_1&=&\half lm\cosh\beta\cosh\gamma\nonumber\\
L_2&=&\half lm\cosh\beta\cosh\alpha\ .
\label{eq:MLfixed}
\end{eqnarray}
The resulting supersymmetric interpretation is then that of
Table~\ref{tbl:inter}, with
\begin{eqnarray}
&&Q_R=Q_L=-2\sqrt{2} M_1\nonumber\\
&&\mu_R=d_L=-2\sqrt{2}L_1
\label{eq:intQ}
\end{eqnarray}
for the electric state, and
\begin{eqnarray}
&&P_R=-P_L=-2\sqrt{2} M_2\nonumber\\
&&d_R=-\mu_L=2\sqrt{2}L_2
\label{eq:intP}
\end{eqnarray}
for the magnetic state.  Note that by looking at an explicit solution, the
left-sided dipole moments are completely determined.  Furthermore, we see
that in fact $\mu_L$ and $d_L$ are more naturally connected to the
magnetic and electric pieces respectively of the intermediate state ansatz.

The heterotic fields may now be translated to the Type II picture using the
duality dictionary given in Eqn.~(\ref{eq:dualitymap}).  The resulting Type
II charges are given in Table~\ref{tbl:interbh}, where $\Delta L\equiv
L_1-L_2$.  Examination of the table indicates a surprising and interesting
result: despite the freedom in taking the supersymmetric limit of
(\ref{classbh}), expressed by the conditions (\ref{eq:MLfixed}) (or,
equivalently, the supersymmetry ambiguity expressed in the
functions $X$ and $Y$), this arbitrariness completely vanishes from the
magnetic dipole moments in the Type II picture.  In fact, for this black
hole, we find $\tilde Q_R=-2\sqrt{2} M$, hence saturating the
Bogomol'nyi bound for $\tilde N_R=1/2$.  The resulting magnetic moment is
purely right-sided, with $\tilde\mu_R=-2\sqrt{2}L=(\tilde Q_R/M)L$,
leading to the gyromagnetic ratios $\tilde g=(0,2)$, in agreement with the
elementary Type II string picture.  Since the $X$ and $Y$ ambiguity resides
solely in the Type II electric dipole moments, we again anticipate a string
calculation of electric dipole moments to complete the picture.  However,
since $\tilde d_L=-2\sqrt{2}\Delta L$, this suggests $\Delta L\approx 0$
if the assumption that such electric dipole moments originate from the Ramond
sector of the string is to be correct.  In turn this indicates the necessity
of setting $\alpha=\gamma$ in Eqn.~(\ref{classbh}), which ensures that
magnetic charges in the Type II picture vanish identically before taking
the limit.

\begin{table}[t]
\begin{center}
\begin{tabular}{c|cccc|cccc}
\multicolumn{1}{c}{}&\multicolumn{4}{c}{Heterotic}&
\multicolumn{4}{c}{Type II}\\
&$F_1$&$F_2$&$F_3$&$F_4$&$\tilde F_1$&$\tilde F_2$&$\tilde F_3$&$\tilde F_4$\\
\hline
\hline
$Q$&$q$&0&0&0&$q$&0&$p$&0\\
$P$&0&0&0&$p$&0&0&0&0\\
\hline
$\mu$&$-2L$&0&$-2\Delta L$&0&$-2L$&0&$-2L$&0\\
$d$&0&$-2\Delta L$&0&$2L$&0&$-2\Delta L$&0&$2\Delta L$\\
\end{tabular}
\end{center}
\caption{Charges and dipole moments of the dyonic black hole in the $L\gg1$
limit. Note that $q$ and $p$ have the same sign.}
\label{tbl:interbh}
\end{table}

To complete the picture for intermediate multiplets, we must now consider
the fermion zero modes.  Starting in the heterotic picture, we recall from
(\ref{eq:intQ}) and (\ref{eq:intP}) that the charges of the black hole are
\begin{eqnarray}
&&Q_R=Q_L=-2\sqrt{2}M_1\nonumber\\
&&P_R=-P_L=-2\sqrt{2}M_2\ .
\end{eqnarray}
Translated to the Type II picture, this corresponds to a purely electric
state with
\begin{eqnarray}
\tilde Q_R&=&-2\sqrt{2}M\nonumber\\
\tilde Q_L&=&-2\sqrt{2}(M_1-M_2)\ .
\end{eqnarray}
The basic zero mode algebra for the intermediate state is somewhat more
complicated than that for short multiplets.  Given the general spinor
$\epsilon$, a quarter of the components correspond to Killing spinors.  The
remaining components are split as one quarter for the electric state ($\vec
S_1$), one quarter for the magnetic state ($\vec S_2$), and finally one
quarter active for both.  Note that, depending on $M_1$ and $M_2$, the zero
modes in general have to be normalized independently, so that the total
spin, $\vec S=\vec S_1+\vec S_2$, is appropriately quantized in basic units
of $1/2$.  Denoting the representation generating creation operators (built
out of the zero modes) as $Q_1^\dagger,Q_2^\dagger,\ldots,Q_6^\dagger$, and
focusing on spin along the $z$-axis, we find the quantities listed in
Table~\ref{tbl:intercreate}, where $\xi=M_1/(M_1+M_2)$ represents the
splitting of mass $M$ into $M_1$ and $M_2$.  The Type II dipole moments have
been calculated using the map
\begin{eqnarray}
\tilde\mu_{L,R}&=&\half[(\mu_R+\mu_L)\pm(d_R-d_L)]\nonumber\\
\tilde d_{L,R}&=&\half[\pm(\mu_R-\mu_L)+(d_R+d_L)]\ ,
\end{eqnarray}
which follows from Eqn.~(\ref{eq:dualitymap}).

\begin{table}[t]
\begin{center}
\begin{tabular}{c|cccccc}
&$Q_1^\dagger$&$Q_2^\dagger$&$Q_3^\dagger$
&$Q_4^\dagger$&$Q_5^\dagger$&$Q_6^\dagger$\\
\hline\hline
$S_1$&$\half$&$-\half$&$\half \xi$&$-\half \xi$&0&0\\
$S_2$&0&0&$\half(1-\xi)$&$-\half(1-\xi)$&$\half$&$-\half$\\
$S=S_1+S_2$&$\half$&$-\half$&$\half$&$-\half$&$\half$&$-\half$\\
\hline
$\mu_L$&$-\sqrt{2}$&$\sqrt{2}$&$-\sqrt{2}\xi$&$\sqrt{2}\xi$&0&0\\
$\mu_R$&0&0&$\sqrt{2}(1-\xi)$&$-\sqrt{2}(1-\xi)$&$-\sqrt{2}$&$\sqrt{2}$\\
$d_L$&0&0&$-\sqrt{2}(1-\xi)$&$\sqrt{2}(1-\xi)$&$-\sqrt{2}$&$\sqrt{2}$\\
$d_R$&$\sqrt{2}$&$-\sqrt{2}$&$-\sqrt{2}\xi$&$\sqrt{2}\xi$&0&0\\
\hline
$\tilde\mu_L$&0&0&$\sqrt{2}(1-2\xi)$&$-\sqrt{2}(1-2\xi)$&0&0\\
$\tilde\mu_R$&$-\sqrt{2}$&$\sqrt{2}$&0&0&$-\sqrt{2}$&$\sqrt{2}$\\
$\tilde d_L$&$\sqrt{2}$&$-\sqrt{2}$&0&0&$-\sqrt{2}$&$\sqrt{2}$\\
$\tilde d_R$&0&0&$-\sqrt{2}$&$\sqrt{2}$&0&0\\
\hline
$(\tilde g_L,\tilde g_R)$&$(0,2)$&$(0,2)$&$(2,0)$&$(2,0)$&$(0,2)$&$(0,2)$
\end{tabular}
\end{center}
\caption{Intermediate state generating creation operators and their
properties in both the heterotic and Type II pictures.  Of the six
operators, $Q_1^\dagger$, $Q_3^\dagger$ and $Q_5^\dagger$ raise the
spin by $1/2$, and $Q_2^\dagger$, $Q_4^\dagger$ and $Q_6^\dagger$ lower the
spin by $1/2$.  Note that $Q_3^\dagger$ and $Q_4^\dagger$ have special
significance as zero modes that are active under both electric and magnetic
projections.}
\label{tbl:intercreate}
\end{table}

In the last line of Table~\ref{tbl:intercreate} we have listed the
gyromagnetic ratios corresponding to the particular zero mode.  This is only
meant to be schematic, as in general several zero modes may be
simultaneously active, and furthermore the dipole moments need to be
combined with those of the bosonic state, given in Table~\ref{tbl:interbh}.
Nevertheless, we see that the non-overlapping
operators, $Q_1^\dagger$, $Q_2^\dagger$, $Q_5^\dagger$, and $Q_6^\dagger$
all have $\tilde g=(0,2)$, while the overlapping operators have $\tilde
g=(2,0)$, hinting at a left-right splitting on the Type II world sheet.
Avoiding the complication of adding Kerr angular momentum to the
zero-mode-generated angular momentum, we now specialize to the superspin
zero multiplet.  In this case it is not hard to see that the four creation
operators corresponding
to $\tilde g=(0,2)$ generate 16 states with spins $[(1)+4(\half)+5(0)]$,
while the other two generate 4 states with spins $[(\half)+2(0)]$.  This
corresponds {\it precisely} to the expected gyromagnetic ratios of the $L=0$
intermediate multiplet generated from the elementary Type II string, with
$\tilde N_R=1/2$, and the left-right splitting of spins on the worldsheet,
$[(1)+4(\half)+5(0)]_L\times[(\half)+2(0)]_R$.

We have now seen that the
gyromagnetic ratios of black holes do indeed correspond in the expected
manner to that of elementary Type II states.  Namely for $\tilde N_R=1/2$
states, the Kerr angular momentum is generated from the left-side of the
Type II string, giving $\tilde g=(0,2)$, whereas the fermion zero modes
filling out the intermediate representation correspond to both left- and
right-sided spin on the Type II worldsheet.  Thus we are left with a most
pleasing conclusion. On one side, the black hole solutions under
investigation have the properties of elementary Type II strings. On the
other hand, this result was not at all enforced by supersymmetry.  Eventually,
the ambiguity left by supersymmetry does not really express itself in the
$X$-$Y$ split, but rather in the disconnected left- and right-sided sector
and the resulting freedom in $\vec\mu^{(0)}_L$ and $\vec d^{(0)}_L$.

\subsection{Non-Supersymmetric States}

It was recently conjectured \cite{Rahmfeld3} that some
non-supersymmetric but nevertheless extremal black holes might be
identified with non-BPS states of the heterotic string.  In particular,
since the bosonic solution by itself does not particularly distinguish
between left- and right-sided gauge fields, black holes may be constructed
that are uncharged under the graviphotons, but charged under the left-sided
vector fields, and which satisfy the extremal condition $M^2=Q_L^2/8$
\cite{Garfinkle:1991}.  In this non-supersymmetric case it is no longer
possible to make general statements about the dipole moments carried by the
bosonic solution.  Nevertheless, it turns out that the fermion zero mode
contributions have a very general form for any black hole solution,
independent of the specifics of the solution.  Thus we may extend the above
techniques to the analysis of gyromagnetic ratios of non-supersymmetric black
holes.  This then allows a further comparison of such states with the
heterotic string.

For non-supersymmetric black holes, we are no longer guided by Killing
spinor equations.  However it is still possible to examine the
supersymmetry variations in the asymptotic regime.  In this case,
since the variations are local expressions, it is clear that the
zero-mode-generated quantities, $\delta\delta(\ldots)$, can only depend on
the mass and asymptotic charges of the solution.  Without supersymmetry to
couple various fields together, we define the additional charges $\delta$,
$N$ and ${\cal R}^{Ia}$ according to
\begin{eqnarray}
\hat g_{ij}&\sim&\delta_{ij}(1+{2\delta\over r})\nonumber\\
e^{-\eta}&\sim&1+{2N\over r}\nonumber\\
V_LL\partial_i V_R^T&\sim&{\cal R}{\hat x_i\over r^2}\ ,
\end{eqnarray}
where the metric retains the form (\ref{eq:metric}).  In particular, we have
picked an isotropic form of the asymptotic metric, with spatial components
$g_{ij}\sim -\hat g_{ij}/g_{00}\sim\delta_{ij}(1+2(M+\delta)/r)$.
Using these definitions, as well as the
standard electric and magnetic charges, results in the zero mode spin
\begin{equation}
\vec S=-{1\over4}\overline{\epsilon}\vec\gamma i\gamma^5[
(M_1-{1\over2\sqrt{2}}\gamma^0Q_R\cdot\Gamma)
+(M_2+{1\over2\sqrt{2}}i\gamma^5\gamma^0P_R\cdot\Gamma)]\epsilon\ ,
\end{equation}
as well as the electric and magnetic dipole moments
\begin{eqnarray}
\delta\delta\vec\mu_R^a&=&{1\over\sqrt{2}}\overline{\epsilon}
\gamma_0\vec\gamma i\gamma^5\Gamma^a
[M_2+{1\over2\sqrt{2}}i\gamma^5\gamma^0P_R\cdot\Gamma]\epsilon\nonumber\\
\delta\delta\vec d_R^a&=&{1\over\sqrt{2}}\overline{\epsilon}
\gamma_0\vec\gamma\Gamma^a
[M_1-{1\over2\sqrt{2}}\gamma^0Q_R\cdot\Gamma]\epsilon\nonumber\\
\delta\delta\vec\mu_L&=&-{1\over4}\overline{\epsilon}
\vec\gamma i\gamma^5[Q_L-\sqrt{2}\gamma^0{\cal R}\cdot\Gamma]\epsilon
\nonumber\\
\delta\delta\vec d_L&=&{1\over4}\overline{\epsilon}
\vec\gamma i\gamma^5[P_L-\sqrt{2}i\gamma^5\gamma^0{\cal R}\cdot\Gamma]
\epsilon\ ,
\end{eqnarray}
where $M_1={1\over2}(M+N)$ and $M_2={1\over2}(M-N+\delta)$.  
While these expressions are similar to those for the intermediate multiplet
ansatz, they are nevertheless quite general.  We see that each dipole moment
receives a contribution from two terms which are precisely related only in
the supersymmetric case.  One thing to note is that no dependence on
the classical (Kerr) angular momentum of the black hole appears in the
above expressions.  This is consistent with the supersymmetric black hole
results, where one can simply combine the zero-mode-generated dipole moments
with their classical counterparts.

In order to study the electrically charged non-supersymmetric
$M^2=Q_L^2/8$ black hole, we may set $\beta=0$ and take the limit
$\alpha\to\infty$ in Eqn.~(\ref{classbh}).  The resulting black hole has
charges
\begin{equation}
Q_L=-2\sqrt{2}M\hat\ell\qquad Q_R=-2\sqrt{2}M\hat n\tanh\gamma
\end{equation}
and only a left-sided magnetic dipole moment,
$\mu_L=-2\sqrt{2}L\hat\ell=(Q_L/M)L$.  Here $\hat\ell$ is a 22-component
unit vector labeling the left-sided $U(1)$ that is active.
{}From the zero modes, we find
in particular $\delta\delta\vec\mu_L=-2\sqrt{2}\vec S\hat\ell=(Q_L/M)\vec
S$ as well as $\delta\delta\vec\mu_R=0$, where the spin $\vec S$
created by the zero modes is
\begin{equation}
\vec S=-{1\over4}M\overline{\epsilon}\vec\gamma i\gamma^5
[1+(\tanh\gamma)\gamma^0\hat n\cdot\Gamma]\epsilon\ .
\end{equation}
This results in the especially simple picture that $g=(2,0)$ for {\it all}
states in the long supermultiplet corresponding to this black hole.  For the
electric dipole moments, we find
\begin{eqnarray}
\delta\delta\vec d_R^a&=&{1\over\sqrt{2}}M\overline{\epsilon}
\gamma_0\vec\gamma\Gamma^a[1+(\tanh\gamma)\gamma^0\hat n\cdot\Gamma]
\epsilon\nonumber\\
\delta\delta\vec d_L&=&{1\over\sqrt{2}}M\hat\ell\overline{\epsilon}
\vec\gamma [1+(\tanh\gamma)\gamma^0\hat n\cdot\Gamma]\epsilon\ .
\end{eqnarray}
In particular, taking the limit $\gamma\to\infty$ reproduces the
supersymmetric results for the Kaluza-Klein black hole, where in this case
$\delta\delta\vec d_L=0$ holds on zero-mode spinors satisfying
Eqn.~(\ref{eq:zeromode}).

Based on the $M^2=Q_L^2/8$ extremal condition, for this black hole to
correspond to a heterotic string state, the latter must have $N_L=1$ and
$N_R=(Q_L^2-Q_R^2+1)/2$.  Thus from the elementary string point of view, the
Kerr angular momentum and quantum spin both originate from the right side
of the string, resulting in the gyromagnetic ratios $g=(2,0)$, in perfect
agreement with the black hole calculation.  We also note that the 
black hole entropy
using the stretched horizon approach \cite{Senblack2} agrees with the
degeneracy of elementary string states. 
On the other hand, the
appearance of non-zero $\delta\delta\vec d_L$ would be somewhat of a
surprise because of its likely origin from a left-sided Ramond sector,
which is of course absent in the heterotic string.  However,
this term
actually vanishes for $Q_R=0$, corresponding to $\gamma=0$, since in
fact $(\overline{\epsilon}\vec\gamma\epsilon)=0$ for spinors satisfying a
ten-dimensional Majorana condition.  What this suggests is that only the
non-supersymmetric extremal black holes that are completely uncharged with
respect to the graviphotons \cite{Garfinkle:1991} may possibly be identified
with elementary heterotic states. However, since the stringy origin 
of electric dipole moments is not totally understood yet, it is too early
to reach a definite conclusion.

\section{Conclusion}

The original motivation for examining electric and magnetic dipole moments
of extremal black holes was to provide a further test of the black holes
as elementary string states conjecture.  We have found out, however, that
for short $N=4$ multiplets the dipole moments were completely fixed by
supersymmetry.  Thus in this case the gyromagnetic ratios do not provide a
true test of the conjecture, as the result is guaranteed by
supersymmetry.  Nevertheless, it is reassuring to see that the correct
values of $(g_L,g_R)$ arise from two completely different derivations---the
string formula \cite{Russo:1995} on the world-sheet, and the supersymmetry
approach in space-time.  We have also made clear the connection between
superspin of the black hole multiplet and the spin originating from the
left side of the heterotic string as well as between the zero-mode spin and
the right side of the string.

A somewhat surprising result of the analysis for short multiplets
was the appearance of electric dipole moments for graviphoton couplings to
electric black holes.  Since these electric dipole moments are only present
for the spin $L\pm1/2$ superpartners of the bosonic spin $L$ solution
(assuming the standard electric black hole of \cite{Senblack1}), from
a string point of view their appearance seems to be related to the Ramond
sector of a supersymmetric worldsheet.  Using heterotic/Type II duality, we
find an intricate structure of electric and magnetic dipole moments on both
sides of the duality map consistent with this interpretation.  The prospect
of understanding how electric dipole moments arise from the worldsheet
point of view is currently under investigation.  Until this is completed it
is also difficult to see how black holes with intrinsic electric dipole
moment \cite{Horowitz:1996a} could possibly fit in with string states.

Turning to intermediate multiplets finally allow a test of the strings and
black holes conjecture.  In this case, however, the comparison must be made
to a Type II string, since the elementary heterotic string has no
intermediate states.  Based on supersymmetry, we find a natural
interpretation of an intermediate multiplet black hole as a combination of
separate electric and magnetic states in the heterotic picture, with both
states independently preserving half of the supersymmetries, yet having only
a quarter preserved in common.  Because supersymmetry leaves several dipole
moments undetermined, we examine the properties of known dyonic black hole
solutions.  Mapping over to the Type II picture, we find agreement with
elementary $\tilde N_R=1/2$ (or $\tilde N_L=1/2$) string states.

Finally, we have given further evidence for the possible identification of
some non-supersymmetric extremal black holes with elementary heterotic
states having $N_L=1$.  While not protected by supersymmetry, this result
may nevertheless indicate the presence of some hidden protection
mechanism in M-theory.  However, since this black hole does not
map into an elementary Type II state, little further insight is
gotten from the dual picture.

In this paper we have often combined the separate notions of black holes as
elementary string states and string/string duality.  In doing so, we have
given some further support for both ideas.  Additionally, we have
seen how well both conjectures complement each other and how useful it can
be to study black holes in various dual pictures.  Although most dyonic
black holes cannot be elementary in any point of view, our results are
consistent with black holes as strings whenever its charges allow it to
be elementary in an appropriate picture.  This outcome suggests an M-theory
approach, in which no single black hole interpretation is particularly more
fundamental than the other.

\bigskip

We wish to thank M.~Porrati for enlightening discussions on gyromagnetic
ratios and for clarifying the connection between $g=2$ and tree-level
unitarity.  JTL wishes to thank I.~Giannakis for many useful inputs, and
also acknowledges the hospitality of the Center for Theoretical Physics 
during the visit while part of this work was being done. JR thanks 
S.J.~Rey for useful discussions. He also thanks the Khuri Lab at 
Rockefeller University, where part of this work was being conducted,
for its hospitality.

\newpage




\end{document}